\documentclass[onecolumn,notitlepage,letterpaper,superscriptaddress,nofootinbib,longbibliography]{revtex4-2}

\usepackage[english]{babel}
\usepackage[T1]{fontenc}
\usepackage[utf8]{inputenc}
\usepackage{amsmath,accents}
\usepackage{amsfonts,color}
\usepackage{amssymb,float}
\usepackage{mathptmx}
\usepackage{comment}
\usepackage{mathrsfs}

\usepackage{graphicx}
\usepackage{subfigure}
\usepackage[dvipsnames]{xcolor}

\usepackage[pdftoolbar = true, pdfstartview = FitH, pdfmenubar = true]{hyperref}
\hypersetup{
    colorlinks=true,
    linkcolor=blue,
    filecolor=magenta,      
   citecolor=blue
}

\newcommand{\weff}{\mathrm{w}_{\mathrm{eff}}}

\newcommand{\Veff}{V_{\mathrm{eff}}}
\newcommand{\meff}{m_{\mathrm{eff}}}

\newcommand{\Acal}{\mathcal{A}}
\newcommand{\Bcal}{\mathcal{B}}
\newcommand{\Vcal}{\mathcal{V}}
\newcommand{\MP}{M_{\mathrm{Pl}}}
\newcommand{\DD}{\mathrm{d}}

\begin{document}

\title{Global Portraits of Inflation in Nonsingular Variables}

\author{Laur Järv} 
\email{laur.jarv@ut.ee}
\affiliation{Institute of Physics, University of Tartu, W.\ Ostwaldi 1, 50411 Tartu, Estonia}
\author{Dmitri Kraiko} 
\email{dmitri.kraiko@ut.ee}
\affiliation{Institute of Physics, University of Tartu, W.\ Ostwaldi 1, 50411 Tartu, Estonia}



\begin{abstract}
In the phase space perspective, scalar field slow roll inflation is described by a heteroclinic orbit from a saddle type fixed point to a final attractive point. In many models the saddle point resides in the scalar field asymptotics, and thus for a comprehensive view of the dynamics a global phase portrait is necessary. For this task, in the literature one mostly encounters dynamical variables that either render the initial or the final state singular, thus obscuring the full picture. In this work we construct a hybrid set of variables which allow the depiction of both the initial and final states distinctly in nonsingular manner. To illustrate the method, we apply these variables to portray various interesting types of scalar field inflationary models like metric Higgs inflation, metric Starobinsky inflation, pole inflation, and a nonminimal Palatini model.
\end{abstract}

\maketitle

\section{Introduction}

Inflationary models of the early universe describe a slowly evolving scalar field that fosters a period of accelerated expansion of space and offer a natural explanation to the observed large scale isotropy and homogeneity, spatial flatness, and the absence of exotic relics \cite{Linde:1981mu, Albrecht:1982wi}. Their predictive power really comes to the forefront in the calculations of the spectrum of density perturbations \cite{Mukhanov:1990me, Lyth:1998xn}, which can be matched to the measurements of the cosmic microwave background radiation \cite{Planck:2018vyg,BICEP:2021xfz}. Among the plethora of different inflationary models proposed \cite{Martin:2013tda}, the tightening observational constraints have ruled out the simplest candidates like scalar fields minimally coupled to gravity and endowed with monomial potentials. This motivates the attention towards nonminimally coupled models \cite{Accetta:1985du, Lucchin:1985ip, Bezrukov:2007ep,Jarv:2016sow} which belong to a broader class of theories referred to as scalar-tensor gravity. An extra nuance must be noted here, namely the distinction between metric and Palatini formalisms, which makes no difference in the minimal coupling, but yields slighlty different field equations in the nonminimal case \cite{Bauer:2008zj} and thus expands the possibilities for fundamental model building \cite{Tenkanen:2020dge, Gialamas:2023flv}.

To generate the observed nearly scale-invariant spectrum of perturbations the scalar field has to sustain a nearly exponential expansion of space. The latter is typically achieved in the slow roll regime, which functions as an attractor for a very robust set of initial conditions \cite{Belinsky:1985zd, Linde:1985ub, Liddle:1994dx}. This feature turns out to be rather generic, and can be succinctly explained in terms of the scalar field phase space. Although qualitatively noted already earlier \cite{Belinsky:1985zd,Felder:2002jk,Urena-Lopez:2007zal,UrenaLopez:2011ur}, it was realized only relatively recently by Alho and Uggla that the phenomenon of the inflationary attractor solution owes its properties to being a heteroclinic orbit from a fixed point with saddle-type features to the final attractor in the phase space \cite{Alho:2014fha, Alho:2017opd}. Such orbits originate from a primordial state with de Sitter effective barotropic index and are often located in the scalar field asymptotics \cite{Alvarez:2019vbp, Quiros:2020bcg, Hrycyna:2020jmw, Jarv:2021qpp, Jarv:2021ehj, Hrycyna:2021yad, Hrycyna:2022der, Alho:2023pkl,Jarv:2024krk}. This state is rather special, as almost all other solutions begin from a different asymptotic state characterized by a power law expansion \cite{Felder:2002jk, Carloni:2007eu, Sami:2012uh, Skugoreva:2014gka, Jarv:2021qpp}, and are drawn to the inflationary orbit only later. Therefore, for a complete description of an inflationary model it is necessary to include the asymptotics and look at the \textit{global} portrait of the phase space.

The methods of phase space analysis have found prolific use in the study of the late universe dominated by dark energy \cite{Bahamonde:2017ize}, but were also applied in the study of inflationary scalar fields from early on \cite{Belinsky:1985zd, Halliwell:1986ja, Amendola:1990nn, Burd:1991ns, Capozziello:1993xn, Kolitch:1994qa, Cornish:1995mf, Copeland:1997et, deOliveira:1997jt, Holden:1998qg, Gunzig:2000kk, Gunzig:2000yj, Gunzig:2000ce, Saa:2000ik, Heard:2002dr, Felder:2002jk, Carloni:2007eu, Abdelwahab:2007jp, Jarv:2008eb, Hrycyna:2008gk, Szydlowski:2013sma, Hrycyna:2013yia, Remmen:2013eja, Arefeva:2012sqa, Skugoreva:2014gka}. For a single scalar field, both minimally or nonminimally coupled, the physical phase space of flat Friedmann--Lema\^itre--Robertson--Walker (FLRW) cosmology is two-dimensional, because the Friedmann constraint relates the Hubble function $H$ to the scalar field $\Phi$ and its time derivative $\dot{\Phi}$. There are several options available how to choose the two dynamical variables that span the physical phase space, as well as how to define ``time'' that parametrizes the evolution. Different choices can bring out or obscure different features, and make the physical interpretation more or less direct.
For instance, many authors have preferred to work with the set that could be schematically represented as $(\sqrt{V}/H,\dot{\Phi}/H)$ \cite{Copeland:1997et}, which has a nice interpretation in terms of the potential and kinetic energy densities of the scalar field, and has been applied in a number of studies \cite{Szydlowski:2008in,Hrycyna:2010yv,Szydlowski:2013sma,Hrycyna:2015eta,UrenaLopez:2011ur,Urena-Lopez:2015odd,Alho:2017opd,Hrycyna:2020jmw,Khyllep:2021yyp}. However, in this approach the dynamical system does not immediately close for an arbitrary potential $V(\Phi)$, and one typically needs to resort to supplementary quantities or variables which complicates the situation. The same can be said about more involved combinations of energy densities \cite{Alho:2017opd,Alho:2022ror,Alho:2023pkl}.
Another approach is to work directly on the $(\Phi, \dot \Phi)$ plane \cite{Grain:2017dqa,Remmen:2013eja,Hergt:2018crm,Amendola:1990nn,Hrycyna:2008gk,Jarv:2008eb,Jarv:2010zc,Arefeva:2012sqa,Skugoreva:2014gka,Pozdeeva:2016cja,Kerachian:2019tar,Mishra:2019ymr,Tenkanen:2020cvw} which can be made compact by using the Poincar\'e transformation. However, in these variables the system becomes singular in the asymptotics and different asymptotic states get mapped to the same point \cite{Jarv:2021qpp}, not ideal for the global analysis. 
An alternative is to take the set $(\Phi, \dot{\Phi}/H)$ \cite{Dutta:2020uha,Jarv:2021ehj,Jarv:2021qpp,Jarv:2024krk} instead, which in yields a regular system and distinct fixed points in the asymptotics, but at the cost of rendering the system singular at the final state where $\dot{\Phi}$, $V(\Phi)$, and $H$ vanish. A further interesting choice is to use $(1/\Phi,\dot{\Phi}/H\Phi)$ \cite{Hrycyna:2015eta,Hrycyna:2020jmw,Hrycyna:2021yad,Hrycyna:2022der}, which turns the usual phase space ``inside out'', i.e.\ maps the scalar field asymptotics to the interior and the final finite state to the outer rim of the compactified phase diagram.

The goal of the present work is to find an optimal set of dynamical variables to span the global phase space of single scalar field FLRW cosmology in scalar-tensor gravity that together with a suitable measure of time would provide an optimal representation of the dynamics of the system, satisfying the following requirements. First, the system must remain regular in the whole phase space including the asymptotics, and map physically different states to distinct points. Second, the variables should work for all models at least in the scalar-tensor family of theories, so that the respective phase portraits of different models can be easily compared next to each other. Third, the portraits should be intuitively easy to read, in the sense that the infinite values of the scalar field are depicted by the outer boundary of the diagams.

We have structured the paper as follows. In Sec.\ \ref{sec:scaltens} we review scalar-tensor gravity and cosmology in metric and Palatini formalisms. Then Sec.\ \ref{sec:dynsys} outlines some general considerations about the choice of the variables and the evolution parameter when presenting single scalar field inflation as a global dynamical system. In Sec.\ \ref{sec:singularvariables} we illustrate how the common variable choices introduce singularities, by taking nonminimally coupled metric Higgs model as an example. The next Sec.\ \ref{sec:quest} continues with the Higgs model and discusses several options how to resolve the issues, finally leading to the construction of ``hybrid'' variables and ``hybrid'' time that satisfy all the requirements. In Sec.\ \ref{sec:examples} we demonstrate that these variables are indeed suitable in the case of other models as well, looking at the examples of metric Starobinsky inflation, pole inflation, and a nonminimal Palatini model with a specific potential, allowing direct comparisons with the diagrams in Ref.\  \cite{Jarv:2024krk}. We end the paper with conclusions and outlook in Sec.\ \ref{sec:conclusion}.

\section{Scalar-tensor cosmology in metric and Palatini formalisms}
\label{sec:scaltens}

A generic action functional for a class of theories which include scalar self-interaction only through the scalar field but not its derivatives can be written as follows \cite{Flanagan:2004bz,Jarv:2014hma,Jarv:2015kga,Jarv:2016sow,Kozak:2018vlp}:\footnote{A more general Palatini action can also contain additional terms due to the nonmetricity of the connection \cite{Kozak:2018vlp}. However, it is possible to show that these extra terms can be eliminated by a conformal transformation of the independent connection, see Appendix A of Ref. \cite{Jarv:2020qqm}.}
\begin{equation} \label{action:flanagan:1}
S = \frac{1}{2}\int_{V_4} \mathrm{d}^4x\sqrt{-g}\left\lbrace {\mathcal A}(\Phi) {\mathcal{R}} - {\mathcal B}(\Phi)g^{\mu\nu}\partial_\mu\Phi \partial_\nu\Phi - 2 \, \mathcal{V}(\Phi)\right\rbrace 
+ S_{(\mathrm{mat})}\left[g_{\mu\nu},\chi_{(\mathrm{mat})}\right] \,.
\end{equation}
Here the Ricci scalar $\mathcal{R}$ can be taken either in the metric formalism as $\mathcal{R}= R \equiv g^{\mu \nu} R_{\mu\nu}$ formed of the Ricci tensor $R_{\mu\nu}$ which is computed from the Levi-Civita connection $\Gamma^\lambda{}_{\mu\nu}$ of the metric $g_{\mu\nu}$, or in the Palatini formalism as $\mathcal{R}= \hat{R} \equiv g^{\mu \nu} \hat{R}_{\mu\nu}$ formed of the Ricci tensor $\hat{R}_{\mu\nu}$ that is computed from a symmetric affine connection $\hat{\Gamma}^\lambda{}_{\mu\nu}$ that is independent of the metric $g_{\mu\nu}$.
The three model functions $\mathcal{A}$, $\mathcal{B}$, and $\mathcal{V}$ fix a concrete theory by specifying the nonminimal coupling to gravity, kinetic self-coupling, and the potential of the scalar field. To avoid instabilities, we  assume
\begin{align}
\label{eq: assumptions on the model functions}
    \mathcal{A}>0 \,, \qquad \mathcal{V}\geq 0 \,,
\end{align}
and 
\begin{equation} 
\label{eq: no ghost condition}
2\mathcal{A}\mathcal{B} + 3 \, \delta_m \left( \mathcal{A}^\prime\right)^2\,  > 0 \,.
\end{equation}
The latter condition ensures that the scalar field is not a ghost, providing the right sign in the scalar field equation \cite{Jarv:2014hma,Jarv:2015kga,Kozak:2018vlp}, and corresponding to a positive-definite field space metric \cite{Kuusk:2015dda,Hohmann:2016yfd}.

In writing the action \eqref{action:flanagan:1} we have assumed that the matter part, $S_{\mathrm(mat)}$, does not depend on the scalar field or on the independent Palatini connection. This guarantees that the free matter particles move along the geodesics of the metric $g_{\mu\nu}$, and their energy-momentum is covariantly conserved w.r.t.\ the Levi-Civita covariant derivative. It is natural to read the speed of light, as well as the units of length, time, and mass to be fixed in this setup, known as the Jordan frame. However, in this frame the effective gravitational constant varies with the scalar field dynamics according to the nonminimal coupling function $\mathcal{A}>0$. Giving the scalar field $\Phi$ mass dimension $1$ as usual, the model functions $\mathcal{A}$, $\mathcal{B}$, $\mathcal{V}$ are of mass dimension 2, 0, and 4, respectively.

In a spatially homogeneous and isotropic flat ($k=0$) spacetime, given by the  Friedmann-Lema\^itre-Robertson-Walker line element
\begin{equation}
\label{FLRW}
\mathrm{d}s^2 = -\mathrm{d}t^2 + a(t)^2 \mathrm{d} \mathbf{x}^2 \,,
\end{equation}
where $t$ corresponds to the cosmic time, the Palatini connection equation can be solved, and the metric and scalar field equations and can be written as \cite{Jarv:2024krk}
\begin{subequations}
\label{eq:FLRW equations}
\begin{align}
\label{mpf:first:friedmann:1} 
H^2 &= -\frac{\mathcal{A}^\prime}{ \mathcal{A} } H \dot{\Phi} + \frac{ 2\mathcal{A} \mathcal{B} - 3 \delta_P (\mathcal{A}')^2}{ 12\mathcal{A}^2 } \dot{\Phi}^2 
+ \frac{\mathcal{V}}{3\mathcal{A}}  \,, \\
\label{mpf:second:friedmann:2} 
2 \dot{H}  +  3H^2 &= -2\frac{\mathcal{A}^\prime}{\mathcal{A}} H \dot{\Phi} 
- \frac{4 \mathcal{A} \mathcal{A}'' + 2 \mathcal{A} \mathcal{B} - 3 \delta_P (\mathcal{A}')^2 }{4\mathcal{A}^2} \dot{\Phi}^2 
- \frac{\mathcal{A}^\prime}{\mathcal{A}}\ddot{\Phi}
+ \frac{\mathcal{V}}{\mathcal{A}}   \,, \\
\label{mpf:scalar:field:equation:friedmann:1}
\ddot{\Phi} &= -3 H\dot{\Phi} - \frac{\mathcal{A} \mathcal{B}' + \mathcal{A}' \mathcal{B} 
+ 3 \, \delta_m \mathcal{A}^\prime \mathcal{A}'' }{2 \mathcal{A} \mathcal{B} 
+ 3 \, \delta_m (\mathcal{A}^\prime)^2}\dot{\Phi}^2  
- \frac{ 2 \mathcal{A} \mathcal{V}^\prime  - 4 \mathcal{V} \mathcal{A}^\prime }{2 \mathcal{A} \mathcal{B} 
+ 3 \, \delta_m (\mathcal{A}^\prime)^2} \,.
\end{align} 
\end{subequations}
Here the prime denotes a derivative with respect to the scalar field, while the dots mark derivatives with respect to the time variable $t$. The expansion of space is characterized by the Hubble function $H=\tfrac{\dot{a}}{a}$.
The symbols
\begin{align}
\label{eq: metric Palatini delta}
    {\delta}_m &= \left\lbrace \begin{tabular}{ll}
        1 \,, & \textrm{metric} \\
        0 \,, & \textrm{Palatini}
    \end{tabular} \right. 
    \,, \qquad
    {\delta}_P = \left\lbrace \begin{tabular}{ll}
        0 \,, & \textrm{metric} \\
        1 \,, & \textrm{Palatini}
    \end{tabular} \right. 
\end{align}
switch on the two extra terms that appear in the tensor field equation in the Palatini case and the two extra terms that appear in the scalar field equation in the metric case. In short, for general model functions, the difference in the metric and Palatini cases reduces to\footnote{A simple metric-Palatini relationship was first noticed in the Brans-Dicke-type case \cite{Lindstrom:1975ry,Lindstrom:1976pq,vandenBergh:1981,Burton:1997pe}, and can be generalized to more elaborate Horndeski-type of theories as well \cite{Helpin:2019kcq}.}
\begin{align}
    \Bcal_{\rm Palatini} &= \Bcal_{\rm metric} + \frac{3 (\Acal')^2}{2 \Acal} \,.
\end{align}
Thus, at least classically, the Palatini formalism does not introduce a new class of theories compared to the metric formalism, but simply corresponds to a reparametrization of the degree of freedom in the scalar field. However, when motivating or postulating a model, e.g.\ coupling the Standard Model Higgs field nonminimally to curvature, the distinction between the metric and Palatini formulation becomes relevant \cite{Bauer:2008zj}. In order to make the formulae maximally transparent and to facilitate the comparison between the metric and Palatini cases, we will keep the delta symbols \eqref{eq: metric Palatini delta} explicit.

In terms of the phase space, the system \eqref{eq:FLRW equations} is 2-dimensional, since we can algebraically express $H$ from the Friedmann constraint \eqref{mpf:first:friedmann:1} and substitute into \eqref{mpf:scalar:field:equation:friedmann:1} which gives a 2nd order differential equation for the scalar field only. The other Friedmann equation \eqref{mpf:second:friedmann:2} derives from \eqref{mpf:first:friedmann:1} and \eqref{mpf:scalar:field:equation:friedmann:1}. Thus once the dynamics of $\Phi$ is known, the evolution of $H$ is determined by the constraint \eqref{mpf:first:friedmann:1}.
Following Ref.\ \cite{Jarv:2021ehj} we can interpret the first two terms on the r.h.s.\ of \eqref{mpf:scalar:field:equation:friedmann:1} as friction, while the third plays the role of the gradient of the effective potential like a ``force'' divided by the effective ``mass'',
\begin{align}
\label{eq: Veff}
    \Veff &= \frac{\mathcal{V}}{\mathcal{A}^2} \,,\\
\label{eq: m_eff}
    \meff &= \frac{2 \mathcal{A}\mathcal{B} + 3 \delta_m (\mathcal{A'})^2}{2 \mathcal{A}^3} \,.
\end{align}
These two quantities determine the main features of  the scalar field dynamics. The fixed points of the scalar field dynamics, i.e.\ where the scalar field evolution stops, $\ddot{\Phi}$, $\dot{\Phi}$ occur at the extrema of the effective potential or at the values where the effective mass diverges, more precisely
\begin{align}
\label{eq:general_fixed_point_condition}
    \frac{\Veff'}{\meff} \Big|_{FP} &= \frac{ 2 \mathcal{A} \mathcal{V}^\prime  - 4 \mathcal{V} \mathcal{A}^\prime }{2 \mathcal{A} \mathcal{B} + 3 \, \delta_m (\mathcal{A}^\prime)^2} \Big|_{FP} = 0 \,.
\end{align}
Note, that in the present context the effective mass describes the scalar field cosmological dynamics and is not directly related to the mass of the scalar quantum particle. Rather the effective mass is proportional to the one-dimensional analog of the field space metric, which for a single field is reduced to a scalar function \cite{Kuusk:2015dda,Hohmann:2016yfd}. For example, we can understand that given a ``force'' the scalar field evolution can slow down because the field gets more ``massive'' or equivalently because the invariant ``distance'' in the field space increases. 
Curiously, while the effective potentials coincide in the metric and Palatini formalism, the effective mass is different. 
However, this difference amounts to a positive factor only, since all terms in \eqref{eq: m_eff} are positive by definition. As a consequence, the metric and Palatini cases possess the same fixed points with the same properties. 

The rate of expansion can be conveniently measured in terms of the effective barotropic index
\begin{align}
\label{eq:weff}
    \weff &= -1 - \frac{2 \dot{H}}{3 H^2} = -1 - \frac{2 \left[2 \mathcal{A} \mathcal{A}^{\prime} \ddot{\Phi} - 2 H \mathcal{A} \mathcal{A}^{\prime} \dot{\Phi} + \left( 2 \mathcal{A} \mathcal{B} + 2 \mathcal{A} \mathcal{A}^{\prime\prime} - 3 \delta_{P} \left(\mathcal{A}^{\prime}\right)^{2}\right)\dot{\Phi}^{2}\right]}{12 H \mathcal{A} \mathcal{A}^{\prime} \dot{\Phi} - 4 \mathcal{A} \mathcal{V} - \left(2 \mathcal{A} \mathcal{B} - 3 \delta_{P} \left(\mathcal{A}^{\prime}\right)^{2}\right) \dot{\Phi}^{2} } \,.
\end{align}
In the case of minimal coupling ($\mathcal{A}\equiv 1$), $\weff$ varies between $-1$ (occurring in the potential dominated regime where the scalar field derivatives can be neglected over the potential), and $+1$ (occurring in the kinetic regime where the potential can be neglected over the derivatives). In the case of nonminimal coupling, the effective barotropic index can take a wider range of values, including the superaccelerating regime where $\weff<-1$ and $\dot{H}>0$, as well as superstiff regime where $\weff>1$. When the value of $\weff$ is constant, we can integrate the left side of \eqref{eq:weff} to get 
\begin{subequations}    
\label{eq:weff_to_a(t)}
\begin{align}
    H &= \frac{2}{3(1+\weff)(t-t_0)} \,, & a &= a_0 (t-t_0)^{\frac{2}{3(1+\weff)}} \,, & \weff &\neq -1 \,, \label{eq:weff_to_a(t) power law} \\
    H &= H_0 \,, & a &=a_0 \, e^{H_0 t} \,, & \weff &= -1 \label{eq:weff_to_a(t) de Sitter} \,, 
\end{align}
\end{subequations}
where $t_0$ and $a_0$ are integration constants. The scalar field finite fixed points are always de Sitter \eqref{eq:weff_to_a(t) de Sitter} \cite{Burd:1991ns,Gunzig:2000kk,Gunzig:2000ce,Carloni:2007eu,Jarv:2008eb,Jarv:2010zc,Jarv:2014hma}. But for the global portraits it is important to know that the scalar field asymptotic states can correspond to power law expansion \eqref{eq:weff_to_a(t) power law} with different values of $\weff$ \cite{Carloni:2007eu,Jarv:2021qpp,Jarv:2024krk}, a de Sitter state \eqref{eq:weff_to_a(t) de Sitter} \cite{Carloni:2007eu,Sami:2012uh,Jarv:2021qpp}, or also an ``asymptotic de Sitter'' state whereby $H$ and $|\Phi|$ diverge, but $\weff$ as defined in \eqref{eq:weff} is still $-1$ \cite{Skugoreva:2014gka,Jarv:2021qpp}.

To resolve the horizon problem and to generate a nearly scale-invariant spectrum of perturbations that made an imprint on the observations today, the scalar field has to sustain a nearly de Sitter expansion during inflation. This is approximated by the slow roll regime where the time derivatives of the field can be considered to be small. Dropping the $\dot{\Phi}$ term in the Friedmann equation~\eqref{mpf:first:friedmann:1} and analogously neglecting the $\ddot{\Phi}$ and $\dot{\Phi}^2$ terms in the scalar field equation \eqref{mpf:scalar:field:equation:friedmann:1} yields the following slow roll conditions \cite{Jarv:2021qpp,Jarv:2024krk}\footnote{A set of improved conditions was recently proposed in Ref.\ \cite{Pozdeeva:2025ied}.}
\begin{subequations}
\label{eq:slowroll}
    \begin{align}
    \label{eq:slowroll1}
    H^2 &= \frac{ \mathcal{V}}{3  \mathcal{A}} \,, \\
    \label{eq:slowroll2}
    3 H\dot{\Phi} &= \frac{ 2\mathcal{A} \mathcal{V}^\prime  - 4 \mathcal{V} \mathcal{A}^\prime }{2 \mathcal{A} \mathcal{B}
    +  3 \delta_m (\mathcal{A}^\prime)^2}  \,,
\end{align}
\end{subequations}
We kept the $\dot{\Phi}$ term in the second equation \eqref{eq:slowroll2}, since otherwise dropping it would have simply reduced the expression to the de Sitter fixed point condition \eqref{eq:general_fixed_point_condition}. Therefore \eqref{eq:slowroll2} characterizes slow roll as a small deviation from the exact de Sitter behavior. 
Note that the slow roll expression is slightly different in the metric and Palatini formalism. In the metric formalism it matches the Einstein frame slow roll conditions translated into Jordan frame \cite{Chiba:2008ia,Akin:2020mcr,Jarv:2021qpp,Karciauskas:2022jzd}.

From the Friedmann constraint \eqref{mpf:first:friedmann:1} and the definitions \eqref{eq: metric Palatini delta} we can express
\begin{subequations}
\label{eq: H in terms of Phi}
\begin{align}
\label{eq: H in terms of Phi pos}
    H =& -\frac{\dot{\Phi} \mathcal{A}^{\prime}}{2 \mathcal{A}} \pm \sqrt{ \frac{\dot{\Phi}{}^2 (\mathcal{A}^{\prime})^2}{4 \mathcal{A}^2} \left( 1-\delta_P + \frac{2 \mathcal{A}\mathcal{B}}{3 (\mathcal{A}')^2} +\frac{4 \mathcal{A}\mathcal{V}}{\dot{\Phi}{}^2 (\mathcal{A}^{\prime})^2} \right) } \\
\label{eq: H in terms of Phi real}
    =& -\frac{\dot{\Phi} \mathcal{A}^{\prime}}{2 \mathcal{A}} \pm \sqrt{ \frac{\left(2 \mathcal{A} \mathcal{B} + 3 \delta_m \left(\mathcal{A}^{\prime}\right)^{2} \right)\dot{\Phi}{}^2+4 \mathcal{A}\mathcal{V}}{12 \mathcal{A}^2}} \,.
\end{align}
\end{subequations}
Physical situations correspond to real $H$ while in the context of inflation we are interested in expanding universe where $H>0$. In fact, our assumptions \eqref{eq: assumptions on the model functions}, \eqref{eq: no ghost condition} already guarantee that $H$ is real for all values of $\Phi$ and $\dot{\Phi}$, due to \eqref{eq: H in terms of Phi real}. The expression \eqref{eq: H in terms of Phi pos} tells that in the metric case the `$+$' branch corresponds to an expanding universe provided $\mathcal{B}>0$, but in the Palatini case the `$+$' branch may involve a contracting behavior well, for instance when $\dot{\Phi}\mathcal{A}'>0$ and $\mathcal{V}$ is small, i.e.\ during the late phase after inflation where the scalar field oscillates around the minimum of the potential.

\section{Inflation as a dynamical system}
\label{sec:dynsys}

To represent the equations \eqref{eq:FLRW equations} as a dynamical system, the first step is to choose the dynamical variables that span the 2-dimensional phase space. It makes sense define one variable as an invertible function of $\Phi$ only, since this guarantees the resulting dynamical system is closed for any forms of the model functions $\mathcal{A}(\Phi), \mathcal{B}(\Phi), \mathcal{V}(\Phi)$. Let us write schematically
\begin{align}
\label{eq:dynamical variables general}
\phi = \phi(\Phi) \,, \qquad z=z(\Phi, \dot{\Phi}, H) \, .
\end{align}
To avoid complications in the analysis the variables should better be dimensionless, thus quite often one needs to introduce an arbitrary positive constant $M$ of mass dimension one to achieve that. The final dynamical equations will be dimensionless, and the constant $M$ will drop out.

The second step is to define the units in which the evolution of the system is described. This measure of evolution should also better be dimensionless. For instance it can be directly related to the cosmological time,
\begin{align}
\label{eq:dtt and dt}
\frac{\DD}{\DD \bar{t}} &= \frac{1}{M}\frac{\DD}{\DD t} \, ,
\end{align}
where $M$ is again a constant of mass dimension one, so that the derivative itself is dimensionless. Alternatively, one may describe the evolution of the universe in e-folds $N=\ln a$, which are dimensionless by definition. This introduces the derivative of dimensionless e-folds ``time'' as
\begin{align}
\label{eq:dN and dt}
\frac{\DD}{\DD N} = \frac{1}{H}\frac{\DD}{\DD t} \, .
\end{align}
As long as we consider expanding universes, the times \eqref{eq:dtt and dt} and \eqref{eq:dN and dt} run in the same direction and yield the same fixed points. The main difference is that in comparison to the cosmic time the e-folds time will ``tick'' quicker when the universe is expanding fast ($H>1$), and the change in the scalar field value measured w.r.t.\ a unit of e-folds time will be less than w.r.t.\ a unit of the cosmic time. On the other hand, when the relative expansion of the universe is slow ($H<1$) the e-folds time will ``tick'' slower, and correspondingly the change in the scalar field value measured w.r.t.\ a unit of e-folds time will be higher than w.r.t.\ a unit of the cosmic time. 
This difference will be extreme in the limits where $H$ vanishes or diverges according to Eq.\ \eqref{eq: H in terms of Phi}. The former can happen for instance in the late stage of the evolution where the scalar field comes to rest at a value where the potential vanishes. The latter may happen in the asymptotic limit where $\dot{\Phi}$ or the potential $\Vcal$ diverges, or at a pole of the kinetic function $\Bcal$. In other words, when $\dot{\Phi}$ and $H$ both approach zero the derivative $\frac{\DD \Phi}{\DD \bar{t}}$ vanishes but not necessarily $\frac{\DD \Phi}{\DD N}$, and correspondingly when $\dot{\Phi}$ diverges the derivative $\frac{\DD \Phi}{\DD \bar{t}}$ must diverge as well while $\frac{\DD \Phi}{\DD N}$ can remain finite. Thus in such situations choosing between \eqref{eq:dtt and dt} and \eqref{eq:dN and dt} can introduce or remove a singularity in the dynamical system.

Finally, using the definitions of the variables \eqref{eq:dynamical variables general} and the field equations \eqref{eq:FLRW equations} we can write the dimensionless dynamical system as
\begin{subequations}
\label{eq:dynsys general}
\begin{align}
    \phi' &= f_1(\phi, z) \,, \\
    z' &= f_2(\phi,z)
\end{align}
\end{subequations}
where ${}'$ stands for a dimensionless derivative, e.g.\ \eqref{eq:dtt and dt} or \eqref{eq:dN and dt}, and $f_1$, $f_2$ are the functions which encode the dynamics of the original equations \eqref{eq:FLRW equations} in these variables. Similarly, we can express the fixed point condition, slow roll condition, effective barotropic index, and physical phase space condition in terms of $\phi, z$. The fixed points of the system are determined by the condition
\begin{align}
\label{eq: general fixed point condition}
    \phi' &= 0 \,, \qquad z'=0 \,,
\end{align}
i.e.\ are given by the values $(\phi^*, z^*)$ where the functions $f_1$, $f_2$ vanish. 
The stability properties of the fixed points can be assessed by linearizing the system near the fixed point, and finding the eigenvalues and eigenvectors of the characteristic matrix
\begin{align}
\label{eq: characteristic matrix general}
    L &= 
    \left[ \begin{matrix}  
    \frac{\partial f_1}{\partial \phi} & \frac{\partial f_1}{\partial z} \\
    \frac{\partial f_2}{\partial \phi} & \frac{\partial f_2}{\partial z}
    \end{matrix} \right]_{(\phi^*,z^*)} 
\end{align}
An eigendirection is attractive if the real part of the respective eigenvalue is negative, and unstable if the real part of the respective eigenvalue is positive.

In order to properly interpret the results of the dynamical analysis one must understand well the mapping between the original quantities and the dynamical variables. There are some well known examples where the dynamical variables naively introduce an extra stretch of the phase space that is not physical in terms of the original quantities \cite{Alho:2016gzi}, or fail to represent the original dynamics to the full extent by presenting a phase space that is too narrow and misses out some points or directions. It is quite obvious that such anomalies can happen when the mapping introduced by the variables \eqref{eq:dynamical variables general} contain singularities, e.g.\ having in the denominator a quantity that can become zero. In this context, one should be aware that specific choices of the variables might introduce artificial singularities into the system, where $f_1$ or $f_2$ strike infinity by construction. To properly interpret the behavior of the system is is important to understand whether the a singularity in $f_1$ or $f_2$ is corresponds to an actual property of the original system, or is an artefact of the choice of variables. Theoretically it is also possible that the dynamical variables map a singularity in the original system to a regular point in the phase space spanned by these variables. In the case of inflation we are currently interested in, it is an imperative that at least during the inflationary slow roll the variables behave well and do not introduce any singularity, otherwise the interpretation breaks down.

While all values of the original quantities $\Phi$ and $\dot{\Phi}$ corresponded to two real values of $H$, it is not guaranteed that rewriting the constraint \eqref{eq: H in terms of Phi} in terms of the dynamical variables \eqref{eq:dynamical variables general} yields real $H$ for the whole span of the new variables. Obviously, the part of phase space where $\phi$ and $z$ yield complex $H$ is not physical. Another complication stems from the fact that $H$ in \eqref{eq: H in terms of Phi} is double valued, this is the nature of the Friedmann constraint independent of whether it is expressed in terms of $\Phi$ and $\dot{\Phi}$, or $\phi$ and $z$. Therefore, it requires extra attention to keep track how the ``$+$'' and ``$-$'' sheets of $H(\Phi,\dot{\Phi})$ are mapped to the ``$+$'' and ``$-$'' sheets of $H(\phi,z)$. The relationship can become rather tricky when $z$ is defined to involve $H$ itself. In the present study when drawing the global portraits we plotted the ``$+$'' sheet of $H(\phi,z)$, calling physical the region where $H$ is real, matching in value to $H(\Phi,\dot{\Phi})$, and which is contiguous to the slow roll inflationary trajectory. Otherwise, especially in the Palatini case, the technically complete analysis would have introduced extra mathematical details that are irrelevant for the physics of inflation, and do not contribute to the main import of the paper. With these premises and as a consistency check we made sure that the boundary of the physical phase space is a trajectory itself that blocks other trajectories to cross over between the physical and nonphysical parts of the phase space. 

The global properties of the system can be revealed with the help of the Poincar\'e compactification \cite{Bahamonde:2017ize}
\begin{align}
\label{eq: Poincare variables}
\phi_p &= \frac{\phi}{\sqrt{1+\phi^2+z^2}}\,, \qquad
z_p = \frac{z}{\sqrt{1+\phi^2+z^2}} \,.
\end{align}
which maps the infinite range of $\phi, z$ to a disc of unit radius, so that the outer edge of the disc corresponds to infinity.
The inverse relation between the compact variables $\phi_p, z_p$ and the original ones \eqref{eq:dynamical variables general} is
\begin{align}
\label{eq: Poincare variables inverse}
\phi=\frac{\phi_p}{\sqrt{1-\phi_p^2-z_p^2}} \,, \qquad z=\frac{z_p}{\sqrt{1-\phi_p^2-z_p^2}} \,.
\end{align}
Thus, we can rewrite the dynamical system \eqref{eq:dynsys general} as
\begin{subequations}
\label{eq:dynsys:infinite}
\begin{align}
\label{eq:dynsys:infinite:phip}
     \phi_p' =& \frac{\partial \phi_p}{\partial \phi} \phi' + \frac{\partial \phi_p}{\partial z} z' \\
     z_p' =& \frac{\partial z_p}{\partial \phi} \phi' + \frac{\partial z_p}{\partial z} z' 
\end{align}
\end{subequations}
where $\phi'$ and $z'$ are the r.h.s.\ of \eqref{eq:dynsys general} with the variables \eqref{eq: Poincare variables inverse} substituted in. Similarly, it is possible to express the slow roll curves, etc., in terms these new variables, to get a compact global picture of the full phase space including the asymptotic regions. The finite fixed points $(\phi^*,z^*)$ map to finite points $(\phi_p^*,z_p^*)$ in the global space. 

In the global picture, most of the solutions originate (or sometimes terminate) at the fixed points which reside in the asymptotics, i.e.\ on the circle $\phi_p^2+z_p^2=1$. We can find such points by substituting $z_p=\pm \sqrt{1-\phi_p^2}$ into \eqref{eq:dynsys:infinite:phip} and asking the r.h.s.\ to vanish. The resulting equation is an algebraic equation for $\phi_p$ which can have several roots $\phi_p^\star$. In general there is a flow along the asymptotic circle since no trajectory can go ``beyond'' infinity, and only the points $(\phi_p^\star,\pm\sqrt{1-(\phi_p^\star)^2})$ where this flow stops can act as sources or sinks for the regular trajectories inside the circle. If the system has a the boundary to the physical phase space area, then where this boundary reaches the asymptotics, there is an asymptotic fixed point. Indeed, since the boundary of the physical phase space is a trajectory itself, the point where this trajectory reaches the asymptotics must be a fixed point, since it can not go further beyond. 

From the expression of the effective barotropic index \eqref{eq:weff} in terms of the global variables $\phi_p, z_p$ it is possible to determine the expansion law and other properties of the physical state corresponding to the  asymptotic fixed points. In many models studied so far \cite{Jarv:2024krk}, the starting point of the slow roll ``master'' trajectory is an asymptotic de Sitter state and the corresponding fixed point has saddle type features, while a majority of other trajectories start at an asymptotic state of power law expansion, and approach the ``master'' trajectory later on. Thus for a complete picture it is important to distinguish the asymptotic states well.

\section{Examples of singular variables}
\label{sec:singularvariables}

To motivate the discussion with concrete examples, let us first consider two choices of variables used in the literature that lead to a singularity in the description of the early or late evolution, respectively. It is instructive to look at a simple system where the physics is well understood, for example the famous nonminimally coupled Higgs model in the metric formalism \cite{Bezrukov:2007ep} specified by the model functions
 \begin{align}
 \label{eq: Higgs metric model}
    \mathcal{A} &= {\MP}^2 + \xi \Phi^2 \,, \qquad
    \mathcal{B} = 1 \,, \qquad
    \mathcal{V} = \frac{\lambda}{4} \left( \Phi^2 - v^2 \right)^2 \,.
\end{align}
Here the positive constants ${\MP}$ and $v$ are of mass dimension one, while $\xi$ and $\lambda$ are dimensionless. The Higgs field will stabilize at the minimum of its potential, $\Phi_0=v$, and for late low energy universe the Planck mass that comes from the gravitational constant is ${\MP}^2 + \xi \Phi_0^2$. Since the energy scale of electroweak symmetry breaking is much lower than the scale of inflation, we can take the Higgs expectation value $v$ to be zero and assume $\Vcal$ to be simply quartic. In this approximation $\MP$ is the Planck mass in the late universe.

\subsection{Direct scalar field variables in cosmic time}

Probably the most straightforward choice of dynamical variables is just to use the scalar field and its derivative in cosmic time \cite{Amendola:1990nn,Hrycyna:2008gk,Jarv:2008eb,Jarv:2010zc,Arefeva:2012sqa,Skugoreva:2014gka,Pozdeeva:2016cja,Remmen:2013eja,Grain:2017dqa,Hergt:2018crm,Kerachian:2019tar,Mishra:2019ymr,Tenkanen:2020cvw}, appropriately made dimensionless as
\begin{align}
\label{eq: original variables}
    \phi &= \frac{\Phi}{M} \,, \qquad \bar{z} = \frac{\dot{\Phi}}{M^2} \,.
\end{align}
To avoid two distinct mass scales it is convenient to identify $M=\MP$. Applying the derivative \eqref{eq:dtt and dt} on the variables \eqref{eq: original variables}, next substituting in $\ddot{\Phi}$ from \eqref{mpf:scalar:field:equation:friedmann:1}, $\dot{H}$ from \eqref{mpf:second:friedmann:2}, positive $H$ from \eqref{eq: H in terms of Phi}, and finally writing $\Phi, \dot{\Phi}$ in terms of $\phi, \bar{z}$ via \eqref{eq: original variables} gives the dynamical system in cosmic time: 
\begin{subequations}
\label{eq: metric Higgs dyn sys finite}
\begin{align}
    \frac{\DD \phi}{\DD \bar{t}} &= z \,, \\
    \frac{\DD \bar{z}}{\DD \bar{t}} &= - \frac{3 H \bar{z}}{M} - \frac{\phi \left(6 \bar{z}^{2} \xi^{2} + \bar{z}^{2} \xi + \lambda \phi^{2}\right)}{6 \phi^{2} \xi^{2} + \phi^{2} \xi + 1} \,,
\end{align}
where 
\begin{align}
\label{eq: H definition Higgs direct cosmic}
    H = \frac{M \left(-6 \bar{z} \phi \xi + \sqrt{36 \bar{z}^{2} \phi^{2} \xi^{2} + 6 \bar{z}^{2} + 3 \lambda \phi^{4} + 3 \phi^{2} \xi \left(2 \bar{z}^{2} + \lambda \phi^{4}\right)}\right)}{6 \left(\phi^{2} \xi + 1\right)}\,.
\end{align}
\end{subequations}
Note that the system is explicitly dimonsionless since the constant $M$ cancels out. Since the Hubble expression \eqref{eq: H definition Higgs direct cosmic}
is always real and nonnegative, the physical phase space extends over the whole infinite span of the variables $\phi, \bar{z}$. The expressions for the effective barotropic index \eqref{eq:weff} and slow roll \eqref{eq:slowroll} can be similarly obtained in terms of these variables. The system is regular for finite $\phi, \bar{z}$, and according to \eqref{eq: general fixed point condition} there is one fixed point $A$ at the origin. The point is nonhyperbolic, as both eigenvalues of the characteristic matrix \eqref{eq: characteristic matrix general} are zero. The only eigenvector is $(1,0)^T$ which means the solutions approach the point $A$ along the horizontal $\phi$ axis. Therefore we can compute 
\begin{align}
    \weff(A) &= \lim_{\phi \to 0} \lim_{\bar{z} \to 0} \weff = -1-8\xi \,,
\end{align}
c.f.\ Table \ref{tab: Higgs fixed points}.
Overall there are no problems in the local description.

\begin{figure}[t]
	\centering
	\subfigure[]{
		\includegraphics[width=0.46\textwidth]{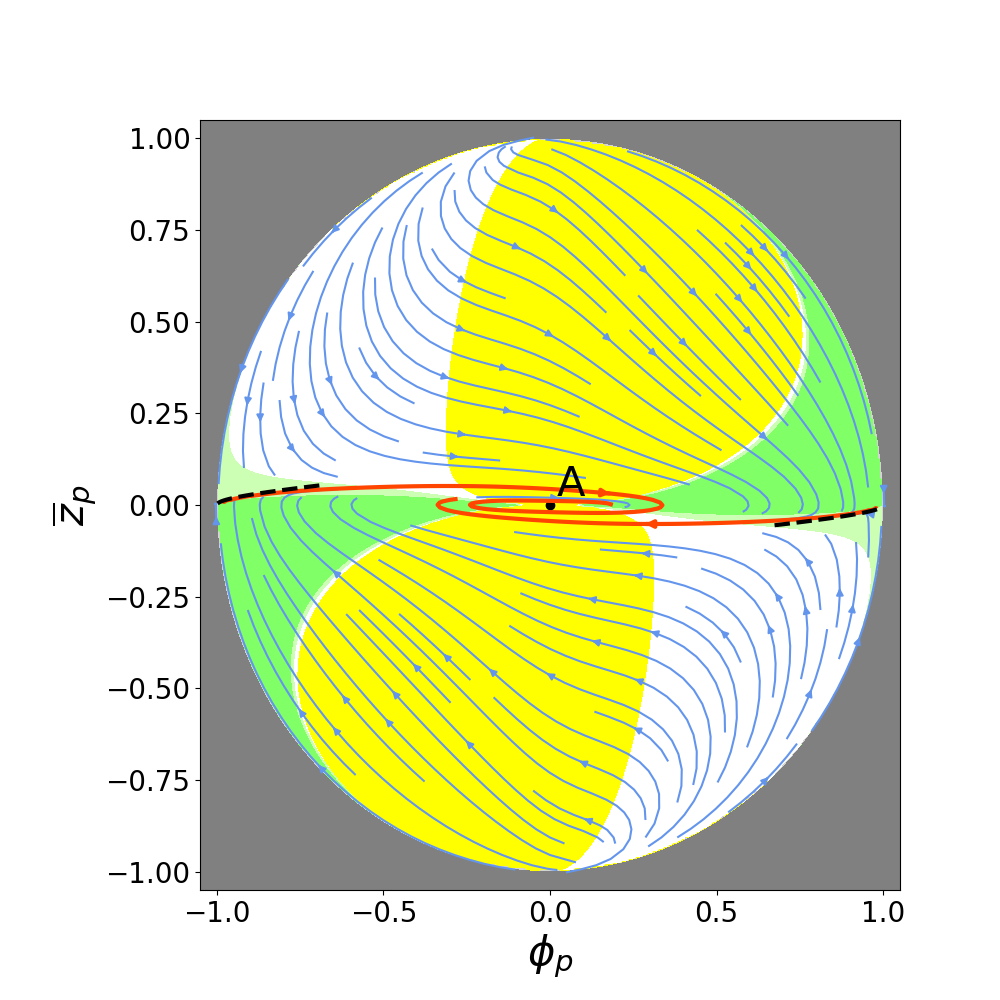} \label{fig: Higgs original variables cosmic xi1}}
    \subfigure[]{
		\includegraphics[width=0.46\textwidth]{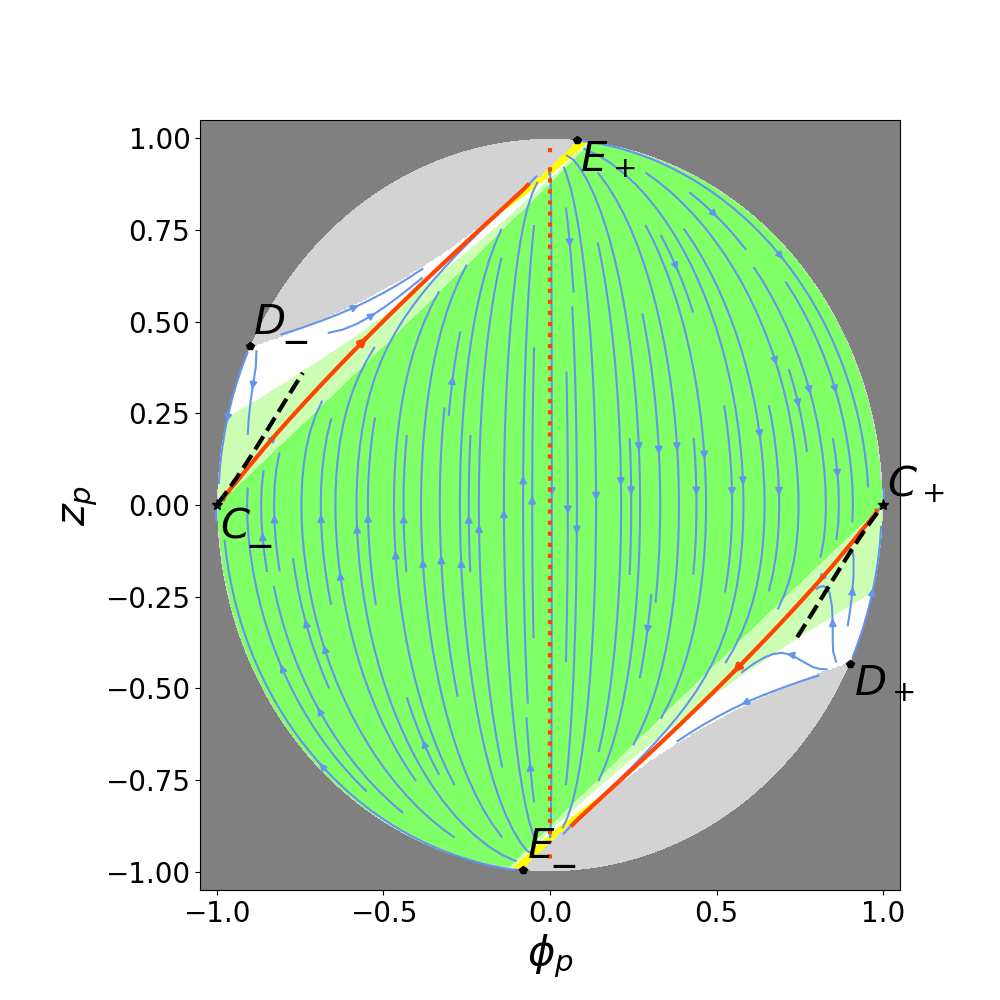} \label{fig: Higgs rescaled variables conformal xi1}}
  \caption{Cosmological phase portraits of the metric Higgs inflationary model model \eqref{eq: Higgs metric model} with $\lambda=0.129$, $v=0$, $\xi=1$ in a) direct scalar field variables \eqref{eq: original variables} in cosmic time \eqref{eq:dtt and dt}, and in b) Hubble rescaled evolution variables \eqref{eq: rescaled variables} in e-folds time \eqref{eq:dN and dt}. The green background stands for superaccelerated, light green accelerated, white decelerated, and yellow superstiff expansion, while grey covers the unphysical region. Orange trajectories are heteroclinic orbits between the fixed points, and the dashed curve marks the path of slow roll approximation. The red dotted line indicates a value where the variables render the system singular.}
\label{fig: singular plots}
\end{figure}

To understand the asymptotic features of the system, we can switch to Poincar\'e compact variables \eqref{eq: Poincare variables}. In terms of these the system \eqref{eq: metric Higgs dyn sys finite} becomes
\begin{subequations}
\label{eq: metric Higgs dyn sys infinite}
\begin{align}
    \frac{\DD \phi_p}{\DD \bar{t}} =& \frac{3 H_p \bar{z}_p^{2} \phi_{p}}{M} + \frac{\bar{z}_p \left(12 \bar{z}_p^{2} \phi_{p}^{2} \xi^{2} + 2 \bar{z}_p^{2} \phi_{p}^{2} \xi + \bar{z}_p^{2} q^{2} + \lambda \phi_{p}^{4} + 6 \phi_{p}^{2} \xi^{2} q^{2} + \phi_{p}^{2} \xi q^{2} + q^{4}\right)}{6 \phi_{p}^{2} \xi^{2} + \phi_{p}^{2} \xi + q^{2}} \,, \\
    \frac{\DD \bar{z}_p}{\DD \bar{t}} =& - \frac{3 H_p \bar{z}_p \left(\phi_{p}^{2} + q^{2}\right)}{M} - \frac{\phi_{p} \left(12 \bar{z}_p^{2} \phi_{p}^{2} \xi^{2} + 2 \bar{z}_p^{2} \phi_{p}^{2} \xi + 6 \bar{z}_p^{2} \xi^{2} q^{2} + \bar{z}_p^{2} \xi q^{2} + \bar{z}_p^{2} q^{2} + \lambda \phi_{p}^{4} + \lambda \phi_{p}^{2} q^{2}\right)}{6 \phi_{p}^{2} \xi^{2} + \phi_{p}^{2} \xi + q^{2}} \,,
\end{align}
where
\begin{align}
\label{eq: H_p definition}
    H_p =& \frac{M \left(-6 \bar{z}_p \phi_{p} \xi q + \sqrt{6 \bar{z}_p^{2} q^{4} + 3 \lambda \phi_{p}^{6} \xi + 3 \phi_{p}^{2} q^{2} \left(12 \bar{z}_p^{2} \xi^{2} + 2 \bar{z}_p^{2} \xi + \lambda \phi_{p}^{2}\right)}\right)}{6 q \left(\phi_{p}^{2} \xi + q^{2}\right)} \,,
\end{align}
\end{subequations}
and the quantity $q^2 = 1- x^2 - y^2$ measures the ``distance'' from the unit circle that represents infinity. The effective barotropic index, slow roll curve, and finite fixed point $A$ can all be mapped onto the unit disc in a regular manner. We can draw the global phase portrait on Fig.\ \ref{fig: Higgs original variables cosmic xi1} (compare with Fig.\ 5 of Ref.\ \cite{Skugoreva:2014gka}). The local dynamics around the point $A$ agrees with the expectations. However, in the asymptotic limit $q \to 0$ the system \eqref{eq: metric Higgs dyn sys finite} diverges. Therefore it is difficult to derive information about the asymptotic dynamics by analytical means. On the figures it is clear that there are asymptotic fixed points on the unit circle, but their precise locations and properties remain elusive. 
Heuristically we can understand the asymptotic divergence of the system already from the equations \eqref{eq: metric Higgs dyn sys finite} where $\frac{d \bar{z}}{d \bar{t}} \sim \phi $ for large $\phi$. Therefore this set of variables is not able to give a fully satisfactory description of the global dynamics.

\subsection{Hubble rescaled variables in e-folds time}

In order to remedy the problems of the direct scalar field variables we can take a second option and like Refs.\ \cite{Jarv:2021ehj,Jarv:2021qpp,Jarv:2024krk} in the definition of the $z$ variable and divide $\dot{\Phi}$ by $H$, i.e.\
\begin{align}
\label{eq: rescaled variables}
    \phi &= \frac{\Phi}{M} \,, \qquad z = \frac{\dot{\Phi}}{HM} \,.
\end{align}
Again, it is natural to identify $M=\MP$. In these variables and using the e-folds time \eqref{eq:dN and dt} the dynamical system of the Higgs model \eqref{eq: H definition Higgs direct cosmic} is given by
\begin{subequations}
\label{eq: metric Higgs dyn sys rescaled finite}
\begin{align}
    \frac{\DD \phi}{\DD N} =& z \,, \\
    \frac{\DD z}{\DD N} =& - \frac{{z} \left(36 \phi^{4} \xi^{3} + 6 \phi^{4} \xi^{2} + 48 \phi^{3} \xi^{3} {z} + 8 \phi^{3} \xi^{2} {z} - 6 \phi^{2} \xi^{2} {z}^{2} + 36 \phi^{2} \xi^{2} - \phi^{2} \xi {z}^{2} + 12 \phi^{2} \xi + 24 \phi \xi^{2} {z} + 8 \phi \xi {z} - 4 \xi {z}^{2} - {z}^{2} + 6\right)}{4 \left(\phi^{2} \xi + 1\right) \left(6 \phi^{2} \xi^{2} + \phi^{2} \xi + 1\right)} \nonumber \\
    & - \frac{M^{2} \lambda \phi^{3} \left(6 \phi^{3} \xi^{2} {z} + \phi^{3} \xi {z} + 8 \phi^{2} \xi + 8 \phi \xi {z} + \phi {z} + 8\right)}{8 H^{2} \left(\phi^{2} \xi + 1\right) \left(6 \phi^{2} \xi^{2} + \phi^{2} \xi + 1\right)} \,,
\end{align}
where
\begin{align}
\label{eq: H in rescaled variables}
    H &= \frac{M \phi^{2}}{2} \sqrt{\frac{2 \lambda}{6 \phi^{2} \xi + 12 \phi \xi {z} - {z}^{2} + 6}} \,.
\end{align}
\end{subequations}
Again note that the system is explicitly dimensionless, as the parameter $M$ cancels out.
The advantage of these variables is that the system \eqref{eq: metric Higgs dyn sys rescaled finite} does not contain any square roots and actually simplifies considerably when $H$ is substituted in. Not all combinations of $\phi, z$ give real $H$, and thus the physically relevant phase space is limited to
\begin{align}
    6 \phi^{2} \xi + 12 \phi \xi {z} - {z}^{2} + 6 \geq 0 \,.
\end{align}
In fact, the bound has a clear physical meaning, as it is exactly satisfied by an extreme kinetic regime of the scalar field dynamics, whereby the derivative terms dominate over the potential, so that the latter can be neglected in the Friedmann equation \eqref{mpf:first:friedmann:1} \cite{Jarv:2021qpp,Jarv:2024krk}. This extreme regime itself is a solution of the equations and thus it blocks the physical solutions from entering into the phase space area that is artificially spanned by the $\phi, z$ variables, but otherwise corresponds to imaginary $H$.

An actual problem with the system \eqref{eq: metric Higgs dyn sys rescaled finite} is that $\frac{dz}{dN}$ becomes singular at $\phi=0$ and we can not properly analyse the fixed point at the origin. Refs.\ \cite{Jarv:2021ehj,Jarv:2021qpp,Jarv:2024krk} resolved this issue by introducing a small cosmological constant $\Lambda$ into the potential $\Vcal$ as a regularizing parameter, computed everything, and in the end took the limit $\Lambda \to 0$. The reasoning supporting these variables is that they faithfully describe the evolution until the end of the accelerated inflationary expansion, and that by the time the scalar field reaches the minimum of the potential, the model ceases to be an accurate description of the physical situation, as reheating and other processes kick in. Therefore a singularity at later times does not matter much for the process of inflation, however it can be an obstacle when following the dynamics into the eras after the inflation. 

In  the global variables \eqref{eq: Poincare variables} the system \eqref{eq: metric Higgs dyn sys rescaled finite} takes a more complicated form 
\begin{subequations}
\label{eq: metric Higgs dyn sys rescaled infinite}
\begin{align}
    \frac{\DD \phi_p}{\DD N} =& \frac{3 \phi_{p}^{4} \xi^{3} z_{p} \left(3 \phi_{p} z_{p} + 2 q^{2} + 6 z_{p}^{2}\right)}{\left(\phi_{p}^{2} \xi + q^{2}\right) \left(6 \phi_{p}^{2} \xi^{2} + \phi_{p}^{2} \xi + q^{2}\right)} \nonumber \\ 
    & + \frac{\phi_{p}^{2} \xi^{2} z_{p} \left(3 \phi_{p}^{3} z_{p} + 2 \phi_{p}^{2} q^{2} + 6 \phi_{p}^{2} z_{p}^{2} + 18 \phi_{p} q^{2} z_{p} - 3 \phi_{p} z_{p}^{3} + 12 q^{4} + 24 q^{2} z_{p}^{2}\right)}{2 \left(\phi_{p}^{2} \xi + q^{2}\right) \left(6 \phi_{p}^{2} \xi^{2} + \phi_{p}^{2} \xi + q^{2}\right)} \nonumber \\
    & + \frac{\phi_{p} \xi z_{p} \left(12 \phi_{p}^{2} q^{2} z_{p} - \phi_{p}^{2} z_{p}^{3} + 8 \phi_{p} q^{4} + 16 \phi_{p} q^{2} z_{p}^{2} - 4 q^{2} z_{p}^{3}\right)}{4 \left(\phi_{p}^{2} \xi + q^{2}\right) \left(6 \phi_{p}^{2} \xi^{2} + \phi_{p}^{2} \xi + q^{2}\right)} + \frac{q^{2} z_{p} \left(6 \phi_{p} q^{2} z_{p} - \phi_{p} z_{p}^{3} + 4 q^{4} + 4 q^{2} z_{p}^{2}\right)}{4 \left(\phi_{p}^{2} \xi + q^{2}\right) \left(6 \phi_{p}^{2} \xi^{2} + \phi_{p}^{2} \xi + q^{2}\right)} \nonumber \\
    & + \frac{M^{2} \lambda \phi_{p}^{4} z_{p} \left(6 \phi_{p}^{3} \xi^{2} z_{p} + \phi_{p}^{3} \xi z_{p} + 8 \phi_{p}^{2} \xi q^{2} + 8 \phi_{p} \xi q^{2} z_{p} + \phi_{p} q^{2} z_{p} + 8 q^{4}\right)}{8 H_p^{2} q^{2} \left(\phi_{p}^{2} \xi + q^{2}\right) \left(6 \phi_{p}^{2} \xi^{2} + \phi_{p}^{2} \xi + q^{2}\right)} \,, \\
    \frac{\DD z_p}{\DD N} =
    & - \frac{3 \phi_{p}^{3} \xi^{3} z_{p} \left(3 \phi_{p}^{3} + 6 \phi_{p}^{2} z_{p} + 3 \phi_{p} q^{2} + 4 q^{2} z_{p}\right)}{\left(\phi_{p}^{2} \xi + q^{2}\right) \left(6 \phi_{p}^{2} \xi^{2} + \phi_{p}^{2} \xi + q^{2}\right)} \nonumber \\
    & - \frac{\phi_{p} \xi^{2} z_{p} \left(3 \phi_{p}^{5} + 6 \phi_{p}^{4} z_{p} + 21 \phi_{p}^{3} q^{2} - 3 \phi_{p}^{3} z_{p}^{2} + 28 \phi_{p}^{2} q^{2} z_{p} + 18 \phi_{p} q^{4} - 3 \phi_{p} q^{2} z_{p}^{2} + 12 q^{4} z_{p}\right)}{2 \left(\phi_{p}^{2} \xi + q^{2}\right) \left(6 \phi_{p}^{2} \xi^{2} + \phi_{p}^{2} \xi + q^{2}\right)} \nonumber \\
    & - \frac{\xi z_{p} \left(12 \phi_{p}^{4} q^{2} - \phi_{p}^{4} z_{p}^{2} + 16 \phi_{p}^{3} q^{2} z_{p} + 12 \phi_{p}^{2} q^{4} - 5 \phi_{p}^{2} q^{2} z_{p}^{2} + 8 \phi_{p} q^{4} z_{p} - 4 q^{4} z_{p}^{2}\right)}{4 \left(\phi_{p}^{2} \xi + q^{2}\right) \left(6 \phi_{p}^{2} \xi^{2} + \phi_{p}^{2} \xi + q^{2}\right)} - \frac{q^{2} z_{p} \left(6 \phi_{p}^{2} q^{2} - \phi_{p}^{2} z_{p}^{2} + 4 \phi_{p} q^{2} z_{p} + 6 q^{4} - q^{2} z_{p}^{2}\right)}{4 \left(\phi_{p}^{2} \xi + q^{2}\right) \left(6 \phi_{p}^{2} \xi^{2} + \phi_{p}^{2} \xi + q^{2}\right)} \nonumber \\
    &- \frac{M^{2} \lambda \phi_{p}^{3} \left(\phi_{p}^{2} + q^{2}\right) \left(6 \phi_{p}^{3} \xi^{2} z_{p} + \phi_{p}^{3} \xi z_{p} + 8 \phi_{p}^{2} \xi q^{2} + 8 \phi_{p} \xi q^{2} z_{p} + \phi_{p} q^{2} z_{p} + 8 q^{4}\right)}{8 H_p^{2} q^{2} \left(\phi_{p}^{2} \xi + q^{2}\right) \left(6 \phi_{p}^{2} \xi^{2} + \phi_{p}^{2} \xi + q^{2}\right)}\,,
\end{align}
where
\begin{align}
\label{eq: H in infinite rescaled variables}
    H_p =& \frac{ M \phi_{p}^{2}}{2q} \sqrt{\frac{2 \lambda}{6 \phi_{p}^{2} \xi + 12 \phi_{p} \xi z_{p} + 6 q^{2} - z_{p}^{2}}} \,.
\end{align}
\end{subequations}
but the pleasant side is that now the system does not diverge in the asymptotics. In fact, we could have anticipated the good asymptotic properties already by looking at \eqref{eq: metric Higgs dyn sys rescaled finite} where $\frac{\DD z}{\DD N} \sim \phi^0$ for large $\phi$. Solving the full system \eqref{eq: metric Higgs dyn sys rescaled infinite} for the fixed points can be cumbersome, but in the process of going to the edge of the unit disc, $q \to 0$, $\phi_p \to \pm\sqrt{1-z_p^2}$ the system reduces to
\begin{subequations}
\label{eq: metric Higgs dyn sys rescaled asymptotic}
\begin{align}
    \frac{d\phi_p}{dN} =& \frac{z_{p}^{2} \left(12 \xi z_{p} \sqrt{1 - z_{p}^{2}} + 6 \xi \left(1 - z_{p}^{2}\right) - z_{p}^{2}\right)}{2 \xi \sqrt{1 - z_{p}^{2}}} \,, \\
    \frac{d z_p}{dN} =& - \frac{z_{p} \left(12 \xi z_{p} \sqrt{1 - z_{p}^{2}} + 6 \xi \left(1 - z_{p}^{2}\right) - z_{p}^{2}\right)}{2 \xi}\,.
\end{align}
\end{subequations}
These two equations are not independent, as we have confined the dynamics to one dimension. Asking the r.h.s.\ of \eqref{eq: metric Higgs dyn sys rescaled asymptotic} to vanish, gives three pairs of asymptotic fixed points $C_\pm$, $D_\pm$, $E_\pm$ (in the notation of Ref.\ \cite{Jarv:2021qpp}) in terms of the quantity 
\begin{align}
    \zeta_\pm &= \frac{\sqrt{\pm 12 \sqrt{6} \xi^{\frac{3}{2}} \sqrt{6 \xi + 1} + 108 \xi^{2} + 6 \xi}}{\sqrt{180 \xi^{2} + 12 \xi + 1}} \,.
\end{align}
These points and their properties are listed in Table \ref{tab: Higgs fixed points}. 

\begin{table}[]
    \centering
    \begin{tabular}{cccc}
         & $(\phi_p^*, z_p^*)$ & type & $\weff$ \\ \hline
       $A$ & $(0,0)$ & nonhyperbolic & $-1-8\xi$ \\
       $C_\pm$ & $\left(\pm 1, 0 \right)$ & nonhyperbolic & $-1$ \\
       $D_\pm$ & $\left(\pm \sqrt{1-\zeta_-^2}, \mp \zeta_- \right)$ & unstable node & $1+8\xi - \frac{4\sqrt{6\xi+36\xi^2}}{3}$ \\
       $E_\pm$ & $\left(\pm \sqrt{1-\zeta_+^2}, \pm \zeta_+ \right)$ & saddle & $1+8\xi + \frac{4\sqrt{6\xi+36\xi^2}}{3}$
    \end{tabular}
    \caption{Fixed points of the nonminimal Higgs model \eqref{eq: Higgs metric model} and their properties \cite{Jarv:2021qpp,Jarv:2024krk}.}
    \label{tab: Higgs fixed points}
\end{table}

The whole phase space is depicted on Fig.\ 
\ref{fig: Higgs rescaled variables conformal xi1}. The points $D_\pm$, $E\pm$ reside at the intersection of the boundary of the physical phase space with the unit circle representing infinity. 
The reason why the variables \eqref{eq: rescaled variables} are doing a better job in distinguishing the asymptotic states than the original variables \eqref{eq: original variables} was discussed in Ref.\ \cite{Jarv:2021qpp}. Namely, in the asymptotic regime the scale factor will typically evolve as $a \sim t^\alpha$ and the scalar field as $\Phi \sim t^\beta$, where $\alpha$ and $\beta$ are some constants. It is easy to check that the ratio $z/\phi \sim {\beta}/{\alpha}$ is finite and maps the asymptotic states to different ``latitudes'' of the unit disc boundary, while the ratio of the original variables, ${\bar{z}}/{\phi}$, will depend on $t$ and map the asymptotic states to either the poles or equator, lumping different regimes together onto a single point. Indeed, by comparing the phase portraits we may recognize how the physically different states represented by the points $D$ and $E$ on Fig.\ \ref{fig: Higgs rescaled variables conformal xi1} are both mapped to the polar points on Fig.\ \ref{fig: Higgs original variables cosmic xi1}. Yet, since the Hubble rescaled evolution variables render the system singular at origin and we can not follow the damped scalar field oscillations  converging to the final state (as was evident with the direct variables in cosmic time), these variables can not be considered to be fully satisfactory either.

\section{Quest for nonsingular variables}
\label{sec:quest}

Having revisited the two sets of variables used in the literature to draw a global phase portrait of Higgs inflation, we will now try to overcome their respective shortcomings, and attempt to construct dynamical variables is such a way that both the initial and final states are represented as distinct fixed points.

\subsection{Direct scalar field variables in e-folds time}
\label{subsec: direct variables conformal time}

Since the system in the direct scalar field variables  \eqref{eq: original variables} and cosmic time \eqref{eq:dtt and dt} diverged in the asymptotics, we may attempt to remedy the situation by keeping these variables and describe the evolution in the e-folds time \eqref{eq:dN and dt} instead. The system now reads 
\begin{subequations}
\label{eq: metric Higgs dyn sys original conformal finite}
\begin{align}
    \frac{\DD \phi}{\DD N} &= \frac{M \bar{z}}{H} \,, \\
    \frac{\DD \bar{z}}{\DD N} &= - 3 \bar{z} - \frac{M \phi \left(6 \bar{z}^{2} \xi^{2} + \bar{z}^{2} \xi + \lambda \phi^{2}\right)}{H \left(6 \phi^{2} \xi^{2} + \phi^{2} \xi + 1\right)} \,,
\end{align}
\end{subequations}
where $H$ is given in \eqref{eq: H definition Higgs direct cosmic}. As noted before, the physical phase space is spanned by the full range of these variables. While in the system \eqref{eq: metric Higgs dyn sys finite} both $\frac{\DD \phi}{\DD \bar{t}}$ and $\frac{\DD \bar{z}}{\DD \bar{t}}$ vanished at the origin, in the e-folds time $\frac{\DD \bar{z}}{\DD N}$ vanishes too but $\frac{\DD \phi}{\DD N}$ is not defined at the origin (we get a different result depending on whether the limit $\phi \to 0$ or $\bar{z} \to 0$ is taken first).  Although it is clear that this combination of the variables and e-folds time does not yield an entirely well behaved description of the system, it is still instructive to check whether the asymptotic properties got improved. At least $\frac{\DD \phi}{\DD N} \sim \phi^{-1}$ and $\frac{\DD \bar{z}}{\DD N} \sim \phi^0$ for large $\phi$, and we can expect a better behavior.

The global equations in terms of the Poincar\'e variables are given by
\begin{subequations}
\label{eq: metric Higgs dyn sys conformal infinite}
\begin{align}
    \frac{\DD \phi_p}{\DD N} =&  3 \bar{z}_p^{2} \phi_{p} + \frac{M \bar{z}_p \left(12 \bar{z}_p^{2} \phi_{p}^{2} \xi^{2} + 2 \bar{z}_p^{2} \phi_{p}^{2} \xi + \bar{z}_p^{2} q^{2} + \lambda \phi_{p}^{4} + 6 \phi_{p}^{2} \xi^{2} q^{2} + \phi_{p}^{2} \xi q^{2} + q^{4}\right)}{H_p \left(6 \phi_{p}^{2} \xi^{2} + \phi_{p}^{2} \xi + q^{2}\right)} \,, \\
    \frac{\DD \bar{z}_p}{\DD N} =& - 3 \bar{z}_p \left(\phi_{p}^{2} + q^{2}\right) - \frac{M \phi_{p} \left(12 \bar{z}_p^{2} \phi_{p}^{2} \xi^{2} + 2 \bar{z}_p^{2} \phi_{p}^{2} \xi + 6 \bar{z}_p^{2} \xi^{2} q^{2} + \bar{z}_p^{2} \xi q^{2} + \bar{z}_p^{2} q^{2} + \lambda \phi_{p}^{4} + \lambda \phi_{p}^{2} q^{2}\right)}{H_p \left(6 \phi_{p}^{2} \xi^{2} + \phi_{p}^{2} \xi + q^{2}\right)} \,,
\end{align}
\end{subequations}
where $H_p$ was given in \eqref{eq: H_p definition}. In contrast to Eqns.\ \eqref{eq: metric Higgs dyn sys infinite} these indeed remain finite in the asymptotic limit. We can explicitly apply the limit $q \to 0$, $\phi_p \to \pm\sqrt{1-\bar{z}_p^2}$ and for nonzero $\xi$ find 
\begin{subequations}
\label{eq: metric Higgs dyn sys direct conformal asymptotic}
\begin{align}
    \frac{\DD \phi_p}{\DD N} =& \pm 3 \bar{z}_p^{2} \sqrt{\left(1- \bar{z}_p \right) \left(1+\bar{z}_p \right)} \,, \\
    \frac{\DD \bar{z}_p}{\DD N} =& -3 \bar{z}_p \left(1-\bar{z}_p \right) \left(1+\bar{z}_p \right) \,.
\end{align}
\end{subequations}
Thus there are two types of fixed points on the asymptotic circle, at $(\pm 1,0)$ and $(0,\pm 1)$. Their characteristic matrices diverge, but we can still identify the first points with $C_\pm$ and the second points with a merger of $D_\pm$ and $E_\pm$ of Table \ref{tab: Higgs fixed points}. In the latter case the effective barotropic index depends on the direction of approach. Indeed, we can compute the limits to find 
\begin{align}
     \lim_{\phi_p \to 0^\pm} \, \lim_{\bar{z}_p\to \mp 1} \weff &= \weff(D_\pm) \,, \\
    \lim_{\phi_p \to 0^\pm} \, \lim_{\bar{z}_p\to \pm 1} \weff &= \weff(E_\pm) \,.
\end{align}
Thus the two physically distinct states are still mapped to the same point in the global phase space.

\begin{figure}[t]
	\centering
	\subfigure[]{
		\includegraphics[width=0.46\textwidth]{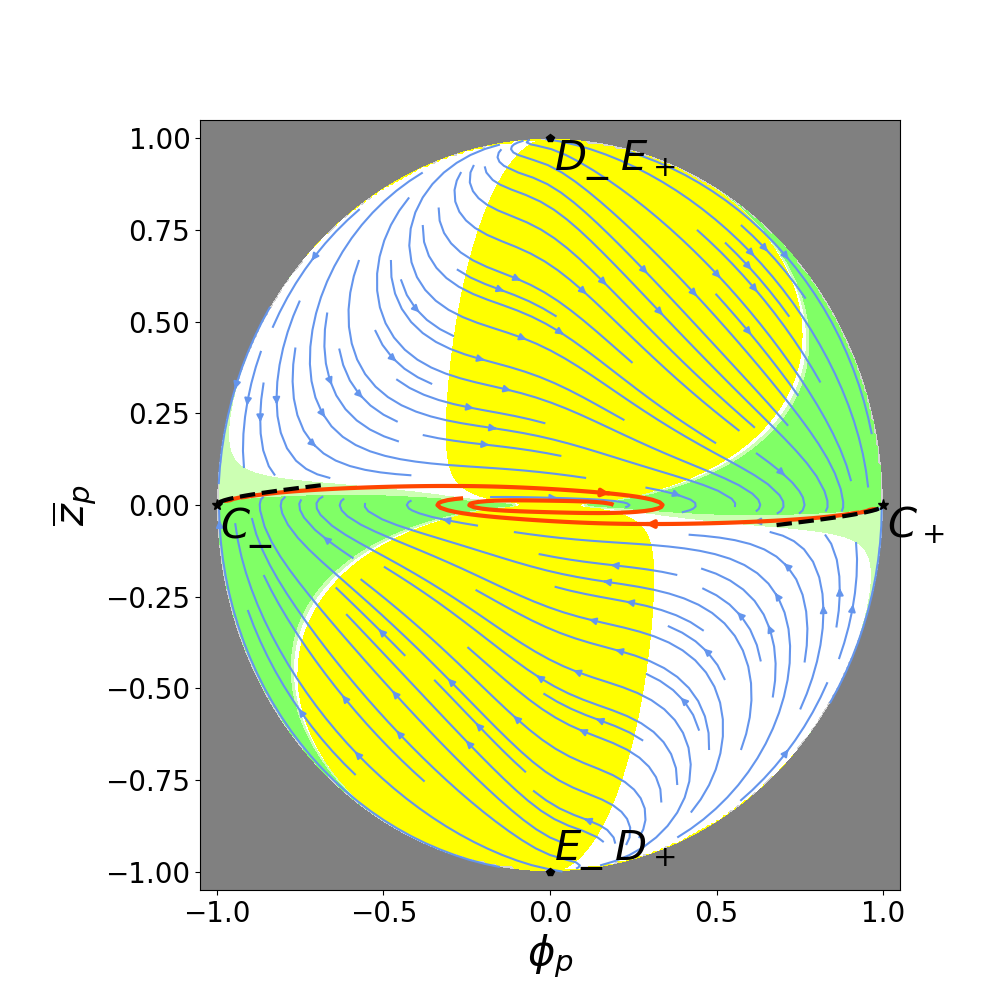} \label{fig: Higgs original variables conformal xi1}}
    \subfigure[]{
		\includegraphics[width=0.46\textwidth]{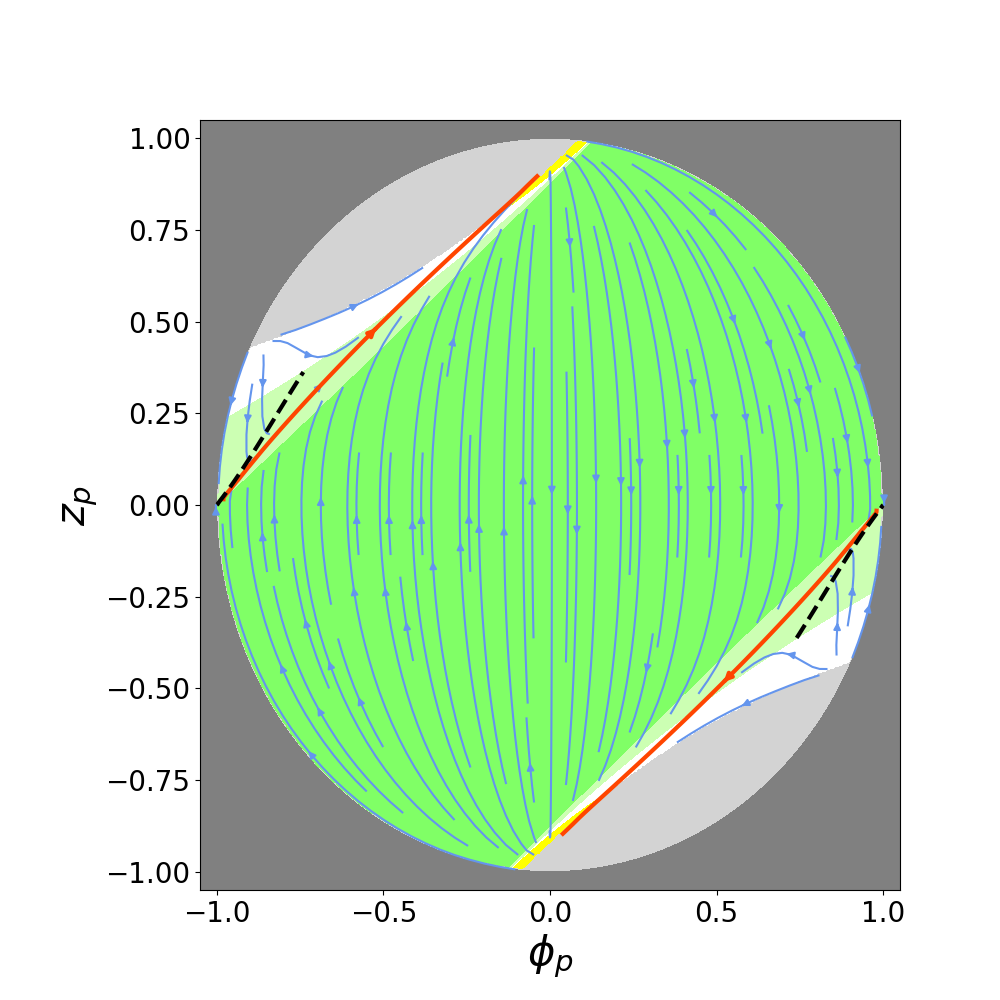} \label{fig: Higgs rescaled variables cosmic xi1}}
  \caption{Cosmological phase portraits of the metric Higgs inflationary model model \eqref{eq: Higgs metric model} with $\lambda=0.129$, $v=0$, $\xi=1$ in a) direct scalar field variables \eqref{eq: original variables} in e-folds time \eqref{eq:dN and dt}, and in b) Hubble rescaled variables \eqref{eq: rescaled variables} in cosmic time \eqref{eq:dtt and dt}. The color coding is the same as on Fig.\ \ref{fig: singular plots}.}
\label{fig: quest for nonsigular plots 2}
\end{figure}

The whole phase space is depicted on Fig.\ \ref{fig: Higgs original variables conformal xi1}. It is qualitatively similar to the phase space pertaining to the same variables in cosmic time, Fig.\ \ref{fig: Higgs original variables cosmic xi1}, only now the asymptotic fixed points can be determined while the central fixed point becomes indiscernible. In principle, one may try to remedy the situation by modifying the e-folds time and make it to depend on $H$, $\Phi$ and $\dot{\Phi}$ \cite{Serna:2002fj,Jarv:2006jd,Jarv:2007iq,Kuusk:2008ak}, but it is not immediately obvious how the problems could be avoided and we will not explore this option any further here.

\subsection{Hubble rescaled variables in cosmic time}

To be systematic, we may also try the Hubble rescaled variables \eqref{eq: rescaled variables} in cosmic time \eqref{eq:dtt and dt}. The system reads now
\begin{subequations}
\label{eq: metric Higgs dyn sys rescaled variables cosmic finite}
\begin{align}
    \frac{\DD \phi}{\DD \bar{t}} =& \frac{Hz}{M} \,, \\
    \frac{\DD z}{\DD \bar{t}} =& 
    \frac{H z \left(- 36 \phi^{4} \xi^{3} - 6 \phi^{4} \xi^{2} - 48 \phi^{3} \xi^{3} z - 8 \phi^{3} \xi^{2} z + 6 \phi^{2} \xi^{2} z^{2} - 36 \phi^{2} \xi^{2} + \phi^{2} \xi z^{2} - 12 \phi^{2} \xi - 24 \phi \xi^{2} z - 8 \phi \xi z + 4 \xi z^{2} + z^{2} - 6\right)}{4 M \left(\phi^{2} \xi + 1\right) \left(6 \phi^{2} \xi^{2} + \phi^{2} \xi + 1\right)} \nonumber \\
    &- \frac{M \lambda \phi^{3} \left(6 \phi^{3} \xi^{2} z + \phi^{3} \xi z + 8 \phi^{2} \xi + 8 \phi \xi z + \phi z + 8\right)}{8 H \left(\phi^{2} \xi + 1\right) \left(6 \phi^{2} \xi^{2} + \phi^{2} \xi + 1\right)} \,,
\end{align}
\end{subequations}
where $H$ is the same as in Eq.\ \eqref{eq: H in rescaled variables} and the physical phase space is limited to the range of $\phi, z$ where $H$ is real. At the value $\phi=0$ the r.h.s.\ of the equations \eqref{eq: metric Higgs dyn sys rescaled variables cosmic finite} vanish, thus the final fixed point $A$ gets ``stretched out'' vertically, and the dynamics around it is obscured.

On the other hand, in the asymptotics the system is singular, and can not provide a satisfactory description of the asymptotic fixed points either. Indeed, while the global equations in terms of the Poincar\'e variables are given by
\begin{subequations}
\label{eq: metric Higgs dyn sys rescaled variables cosmic infinite}
\begin{align}
    \frac{\DD \phi_p}{\DD \bar{t}} =&  \frac{3 H_p \phi_{p}^{4} \xi^{3} z_{p} \left(3 \phi_{p} z_{p} + 2 q^{2} + 6 z_{p}^{2}\right)}{M \left(\phi_{p}^{2} \xi + q^{2}\right) \left(6 \phi_{p}^{2} \xi^{2} + \phi_{p}^{2} \xi + q^{2}\right)} + \frac{H_p \phi_{p}^{2} \xi^{2} z_{p} \left(3 \phi_{p}^{3} z_{p} + 2 \phi_{p}^{2} q^{2} + 6 \phi_{p}^{2} z_{p}^{2} + 18 \phi_{p} q^{2} z_{p} - 3 \phi_{p} z_{p}^{3} + 12 q^{4} + 24 q^{2} z_{p}^{2}\right)}{2 M \left(\phi_{p}^{2} \xi + q^{2}\right) \left(6 \phi_{p}^{2} \xi^{2} + \phi_{p}^{2} \xi + q^{2}\right)} \nonumber \\  
    &+ \frac{H_p \phi_{p} \xi z_{p} \left(12 \phi_{p}^{2} q^{2} z_{p} - \phi_{p}^{2} z_{p}^{3} + 8 \phi_{p} q^{4} + 16 \phi_{p} q^{2} z_{p}^{2} - 4 q^{2} z_{p}^{3}\right)}{4 M \left(\phi_{p}^{2} \xi + q^{2}\right) \left(6 \phi_{p}^{2} \xi^{2} + \phi_{p}^{2} \xi + q^{2}\right)} + \frac{H_p q^{2} z_{p} \left(6 \phi_{p} q^{2} z_{p} - \phi_{p} z_{p}^{3} + 4 q^{4} + 4 q^{2} z_{p}^{2}\right)}{4 M \left(\phi_{p}^{2} \xi + q^{2}\right) \left(6 \phi_{p}^{2} \xi^{2} + \phi_{p}^{2} \xi + q^{2}\right)} \nonumber \\
    &+ \frac{M \lambda \phi_{p}^{4} z_{p} \left(6 \phi_{p}^{3} \xi^{2} z_{p} + \phi_{p}^{3} \xi z_{p} + 8 \phi_{p}^{2} \xi q^{2} + 8 \phi_{p} \xi q^{2} z_{p} + \phi_{p} q^{2} z_{p} + 8 q^{4}\right)}{8 H_p q^{2} \left(\phi_{p}^{2} \xi + q^{2}\right) \left(6 \phi_{p}^{2} \xi^{2} + \phi_{p}^{2} \xi + q^{2}\right)} \,, \\
    \frac{\DD {z}_p}{\DD \bar{t}} =& - \frac{3 H_p \phi_{p}^{3} \xi^{3} z_{p} \left(3 \phi_{p}^{3} + 6 \phi_{p}^{2} z_{p} + 3 \phi_{p} q^{2} + 4 q^{2} z_{p}\right)}{M \left(\phi_{p}^{2} \xi + q^{2}\right) \left(6 \phi_{p}^{2} \xi^{2} + \phi_{p}^{2} \xi + q^{2}\right)} \nonumber \\
    &- \frac{H_p \phi_{p} \xi^{2} z_{p} \left(3 \phi_{p}^{5} + 6 \phi_{p}^{4} z_{p} + 21 \phi_{p}^{3} q^{2} - 3 \phi_{p}^{3} z_{p}^{2} + 28 \phi_{p}^{2} q^{2} z_{p} + 18 \phi_{p} q^{4} - 3 \phi_{p} q^{2} z_{p}^{2} + 12 q^{4} z_{p}\right)}{2 M \left(\phi_{p}^{2} \xi + q^{2}\right) \left(6 \phi_{p}^{2} \xi^{2} + \phi_{p}^{2} \xi + q^{2}\right)} \nonumber \\
    &- \frac{H_p \xi z_{p} \left(12 \phi_{p}^{4} q^{2} - \phi_{p}^{4} z_{p}^{2} + 16 \phi_{p}^{3} q^{2} z_{p} + 12 \phi_{p}^{2} q^{4} - 5 \phi_{p}^{2} q^{2} z_{p}^{2} + 8 \phi_{p} q^{4} z_{p} - 4 q^{4} z_{p}^{2}\right)}{4 M \left(\phi_{p}^{2} \xi + q^{2}\right) \left(6 \phi_{p}^{2} \xi^{2} + \phi_{p}^{2} \xi + q^{2}\right)} \nonumber \\
    &- \frac{H_p q^{2} z_{p} \left(6 \phi_{p}^{2} q^{2} - \phi_{p}^{2} z_{p}^{2} + 4 \phi_{p} q^{2} z_{p} + 6 q^{4} - q^{2} z_{p}^{2}\right)}{4 M \left(\phi_{p}^{2} \xi + q^{2}\right) \left(6 \phi_{p}^{2} \xi^{2} + \phi_{p}^{2} \xi + q^{2}\right)} \nonumber \\
    &- \frac{M \lambda \phi_{p}^{3} \left(\phi_{p}^{2} + q^{2}\right) \left(6 \phi_{p}^{3} \xi^{2} z_{p} + \phi_{p}^{3} \xi z_{p} + 8 \phi_{p}^{2} \xi q^{2} + 8 \phi_{p} \xi q^{2} z_{p} + \phi_{p} q^{2} z_{p} + 8 q^{4}\right)}{8 H_p q^{2} \left(\phi_{p}^{2} \xi + q^{2}\right) \left(6 \phi_{p}^{2} \xi^{2} + \phi_{p}^{2} \xi + q^{2}\right)}\,,
\end{align}
\end{subequations}
where $H_p$ is the same as in \eqref{eq: H in infinite rescaled variables}, this system diverges in the asymptotic limit, in contrast to Eqns.\ \eqref{eq: metric Higgs dyn sys rescaled infinite}. 
Hence we are not able to handle neither the regular nor the asymptotic fixed points. Therefore, this combination of the variables and time choice falls short of our needs for a clear depiction and analyisis of the dynamics of the system. For reference, the whole phase space is depicted on Fig.\ \ref{fig: Higgs rescaled variables cosmic xi1}. Apart from the fixed points, it is qualitatively similar to the phase space pertaining to the same variables in e-folds time, Fig.\ \ref{fig: Higgs rescaled variables conformal xi1}.

\subsection{Hybrid variables in hybrid time}

In order to solve the issue of different variables becoming singular either at the origin (late time fixed point) or in the asymptotics (initial state of the universe) we may try to introduce a set of hybrid variables
\begin{align}
\label{eq: hybrid variables}
    \phi &= \frac{\Phi}{M} \,, \qquad \tilde{z} = \frac{\dot{\Phi}}{(H+M)M} \,.
\end{align}
For small $H$ of the late universe these variables approximate the direct scalar field variables \eqref{eq: original variables} and should avoid becoming singular. In contrast, for large $H$ of the early universe, these variables approximate the Hubble rescaled evolution variables \eqref{eq: rescaled variables} and should keep the system finite in the asymptotic limit. However, the former only happens in cosmic time and the latter only in e-folds time. Therefore in order to reconcile these limits we are obliged to introduce hybrid time 
\begin{align}
\label{eq:dttilde and dt}
\frac{\DD}{\DD \tilde{t}} = \frac{1}{H+M}\frac{\DD}{\DD t} \,
\end{align}
as well. As before, it is natural to identify $M=\MP$.

In the hybrid variables \eqref{eq: hybrid variables} and using the hybrid time \eqref{eq:dttilde and dt} the dynamical system of the Higgs model \eqref{eq: H definition Higgs direct cosmic} is given by
\begin{subequations}
\label{eq: metric Higgs dyn sys hybrid cosmic finite}
\begin{align}
    \frac{\DD \phi}{\DD \tilde{t}} =& \tilde{z} \,, \\
    \frac{\DD \tilde{z}}{\DD \tilde{t}} =& - \frac{H^{2} \tilde{z} \left(36 \phi^{4} \xi^{3} + 6 \phi^{4} \xi^{2} + 48 \phi^{3} \tilde{z} \xi^{3} + 8 \phi^{3} \tilde{z} \xi^{2} - 6 \phi^{2} \tilde{z}^{2} \xi^{2} - \phi^{2} \tilde{z}^{2} \xi + 36 \phi^{2} \xi^{2} + 12 \phi^{2} \xi + 24 \phi \tilde{z} \xi^{2} + 8 \phi \tilde{z} \xi - 4 \tilde{z}^{2} \xi - \tilde{z}^{2} + 6\right)}{4 \left(H + M\right)^2 \left(\phi^{2} \xi + 1\right) \left(6 \phi^{2} \xi^{2} + \phi^{2} \xi + 1\right)} \nonumber \\ 
    & - \frac{H M \tilde{z} \left(36 \phi^{4} \xi^{3} + 6 \phi^{4} \xi^{2} + 36 \phi^{3} \tilde{z} \xi^{3} + 6 \phi^{3} \tilde{z} \xi^{2} - 6 \phi^{2} \tilde{z}^{2} \xi^{2} - \phi^{2} \tilde{z}^{2} \xi + 36 \phi^{2} \xi^{2} + 12 \phi^{2} \xi + 24 \phi \tilde{z} \xi^{2} + 6 \phi \tilde{z} \xi - 4 \tilde{z}^{2} \xi - \tilde{z}^{2} + 6\right)}{2 \left(H + M\right)^2 \left(\phi^{2} \xi + 1\right) \left(6 \phi^{2} \xi^{2} + \phi^{2} \xi + 1\right)} \nonumber \\
    & - \frac{M^{2} \left(6 \lambda \phi^{6} \tilde{z} \xi^{2} + \lambda \phi^{6} \tilde{z} \xi + 8 \lambda \phi^{5} \xi + 8 \lambda \phi^{4} \tilde{z} \xi + \lambda \phi^{4} \tilde{z} + 8 \lambda \phi^{3}\right)}{8 \left(H + M\right)^2 \left(\phi^{2} \xi + 1\right) \left(6 \phi^{2} \xi^{2} + \phi^{2} \xi + 1\right)} \nonumber \\
    & - \frac{M^2 \left(48 \phi^{3} \tilde{z}^{2} \xi^{3} + 8 \phi^{3} \tilde{z}^{2} \xi^{2} - 12 \phi^{2} \tilde{z}^{3} \xi^{2} - 2 \phi^{2} \tilde{z}^{3} \xi + 48 \phi \tilde{z}^{2} \xi^{2} + 8 \phi \tilde{z}^{2} \xi - 8 \tilde{z}^{3} \xi - 2 \tilde{z}^{3}\right)}{8 \left(H + M\right)^2 \left(\phi^{2} \xi + 1\right) \left(6 \phi^{2} \xi^{2} + \phi^{2} \xi + 1\right)}\,,
\end{align}
where
\begin{align}
    H =& \frac{M \left(-12 \phi \tilde{z} \xi + 2 \tilde{z}^{2} + \sqrt{12 \lambda \phi^{6} \xi + 24 \lambda \phi^{5} \tilde{z} \xi - 2 \lambda \phi^{4} \tilde{z}^{2} + 12 \lambda \phi^{4} + 144 \phi^{2} \tilde{z}^{2} \xi^{2} + 24 \phi^{2} \tilde{z}^{2} \xi + 24 \tilde{z}^{2}}\right)}{2 \left(6 \phi^{2} \xi + 12 \phi \tilde{z} \xi - \tilde{z}^{2} + 6\right)} \,.
\end{align}
\end{subequations}
Despite the addition of $M$ and $H$ into the definition of the variables \eqref{eq: hybrid variables} and time \eqref{eq:dttilde and dt} it is easy to check that the parameter $M$ cancels out in the equations and the system is still explicitly dimensionless. One can also check that the fixed point $A$ is present and regular at the origin, with the properties as given in Table \ref{tab: Higgs fixed points}. That feature owes to the fact that the hybrid variables reduce to the direct scalar field variables and the hybrid time reduces to the cosmic time in the small $H$ regime of late universe. On the other hand, the equations are finite in the scalar field asymptotics as $\frac{\DD \tilde{z}}{\DD \tilde{t}} \sim \phi^0$. That feature arises because the hybrid variables reduce to the Hubble rescaled evolution variables and the hybrid time to the e-folds time in the large $H$ regime of the early universe. Thus it is not a miracle that we recover the asymptotic fixed points with the properties given in  Table \ref{tab: Higgs fixed points}. In summary, the dynamical system in the hybrid variables \eqref{eq: hybrid variables} and hybrid time \eqref{eq:dttilde and dt} is globally finite and renders both the initial and final states as regular fixed points, as we have desired.

It is now a straightforward exercise to convert the variables \eqref{eq: hybrid variables} into the  Poincar\'e form \eqref{eq: Poincare variables} and write the system \eqref{eq: metric Higgs dyn sys hybrid cosmic finite} in the global fashion. However, the full expressions get really lengthy and unilluminating, and we skip their explicit presentation here. Nevertheless, with enough effort it is possible to check that the system remains finite both at the origin and in the asymptotics. Utilizing the same method as before, one can confirm the existence and properties of the fixed points as reported in Table \ref{tab: Higgs fixed points} before. Thus the hybrid variables do not introduce extra new features to the system, they only enable the description of these features in a regular and uniform manner.

\begin{figure}[t]
	\centering
	\subfigure[]{
		\includegraphics[width=0.46\textwidth]{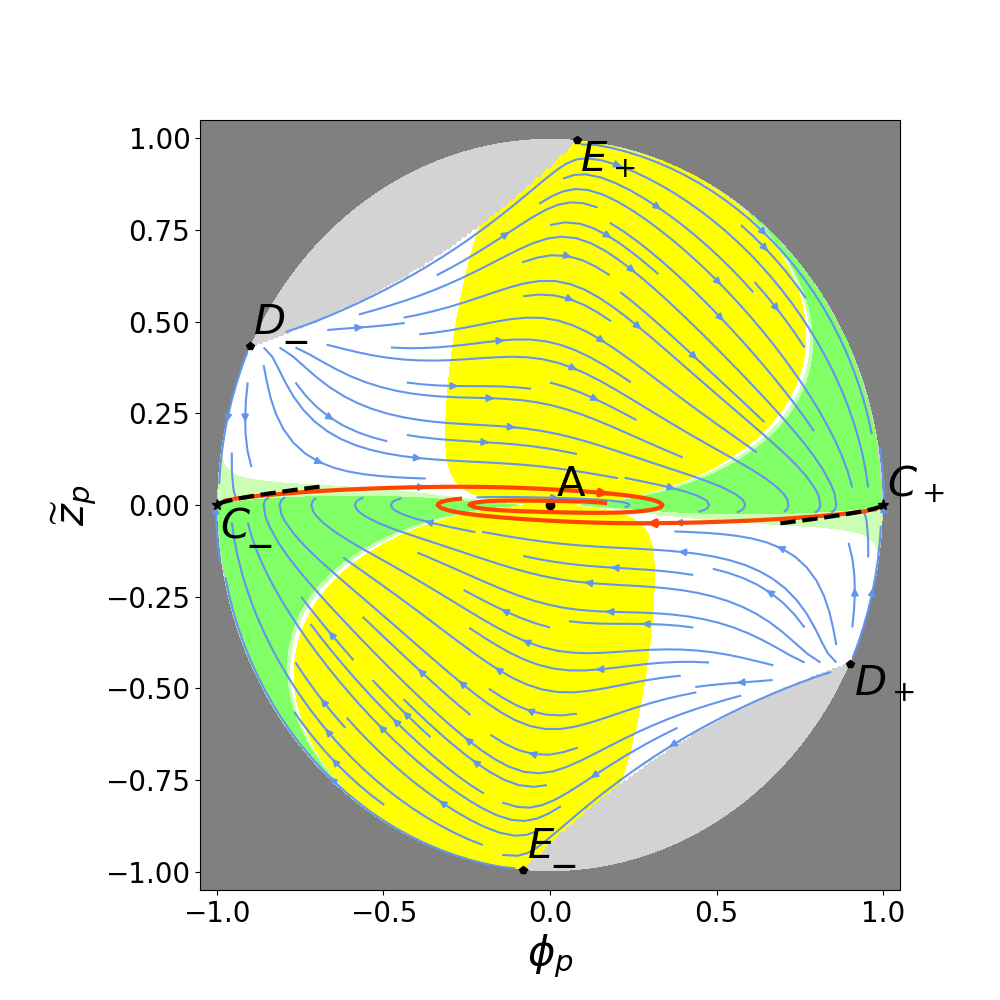} \label{fig: Higgs hybrid variables hybrid time xi1}}
    \subfigure[]{
		\includegraphics[width=0.46\textwidth]{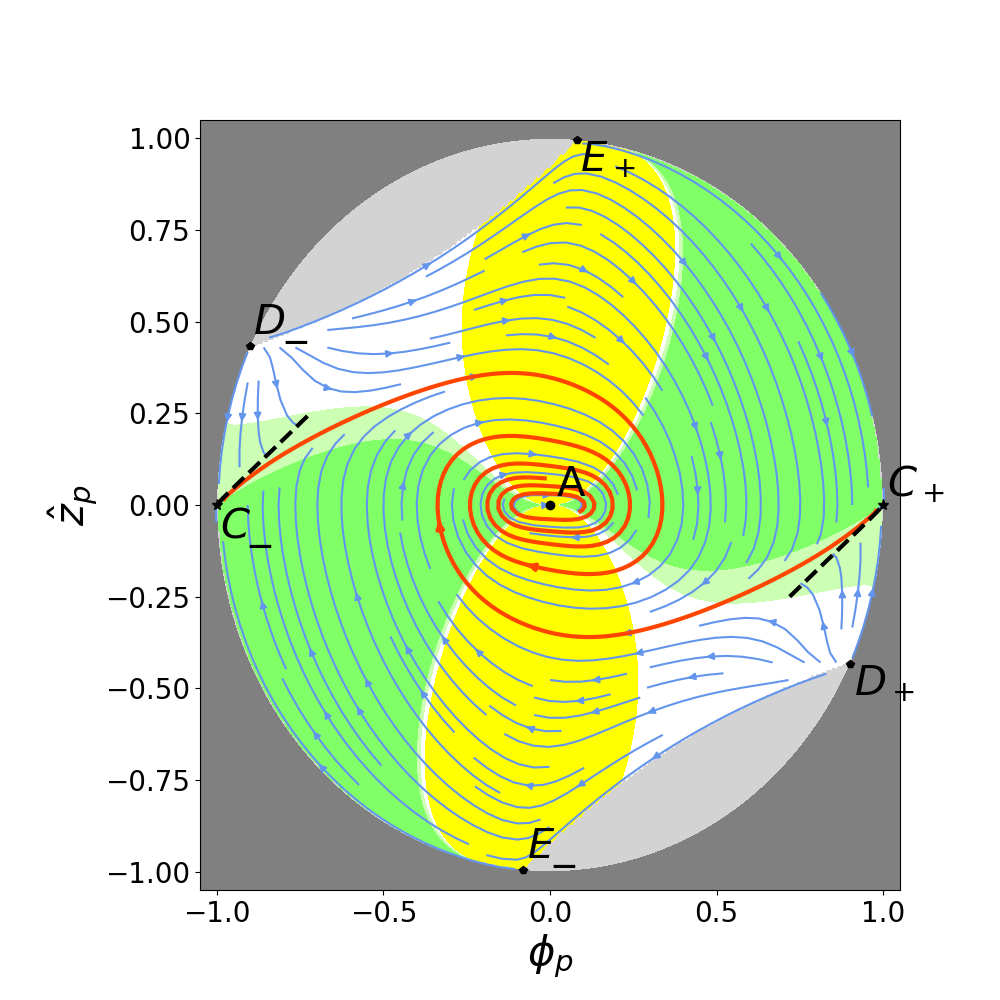} \label{fig: Higgs modified hybrid variables hybrid time xi1}}
  \caption{Cosmological phase portraits of the metric Higgs inflationary model model \eqref{eq: Higgs metric model} with $\lambda=0.129$, $v=0$, $\xi=1$ in a) hybrid variables \eqref{eq: modified hybrid variables} in hybrid time \eqref{eq:dttilde and dt}, and in b) hybrid variables with adjustable scaling factor $s=0.1$ \eqref{eq: modified hybrid variables} in hybrid time \eqref{eq:dtt and dt}. The color coding is the same as on Fig.\ \ref{fig: singular plots}.}
\label{fig: quest for nonsigular plots}
\end{figure}

The global phase portrait of the system is plotted on Fig.\ \ref{fig: Higgs hybrid variables hybrid time xi1}. As expected, at the origin resides the final fixed point $A$ of the late universe, with the features similar to Fig.\ \ref{fig: Higgs original variables cosmic xi1} of the direct variables in cosmic time. On the other hand, in the asymptotics we see the fixed points $C$, $D$, and $E$ with the features similar to Fig.\ \ref{fig: Higgs rescaled variables conformal xi1} of the Hubble rescaled variables in e-folds time. The presence of an unphysical region of the phase space is also inherited from the Hubble rescaled variables, similar to Fig.\ \ref{fig: Higgs rescaled variables conformal xi1}. Overall, visually, the phase portrait \ref{fig: Higgs hybrid variables hybrid time xi1} in the hybrid variables and hybrid time looks like a smooth amalgamation of the inner part of Fig.\ \ref{fig: Higgs original variables cosmic xi1} and the outer part of Fig.\ \ref{fig: Higgs rescaled variables conformal xi1}, combining the desired features of both.

\subsection{Hybrid variables with adjustable rescaling in hybrid time}
\label{subsec: hybrid adjustable rescaled variables}

The hybrid variables in hybrid time resolve the issue of the system becoming singular at the initial or final states. But there is still one minor aspect that can be improved. Namely, in the example plots we saw, the trajectories that spiral into the final state are depicted in a rather squeezed manner. This effect will become even more pronounced for larger $\xi$, making it very hard to obtain a clear visual picture of the actual dynamics. The reason is that in the main part of these final stages of the evolution, at the damped oscillations around the attractor point $A$, the variable $\tilde{z}$ is much smaller than $\phi$. We can gently counter such vertical ``squeezing'' of the trajectories by introducing an extra positive factor $s$ to slightly modify the hybrid variables:
\begin{align}
\label{eq: modified hybrid variables}
    \phi &= \frac{\Phi}{M} \,, \qquad \hat{z} = \frac{\dot{\Phi}}{(H+Ms)M} \,.
\end{align}
For large $H$ of the early universe the $s$ parameter plays no role and leaves the asymptotic initial states as before. For small $H$ of the late universe the $s$ parameter makes the modified hybrid variable $\hat{z}$ bigger than the previous hybrid variable $\tilde{z}$ when $s<1$ and smaller if $s>1$. Therefore depending on the situation the inspiral into the final point $A$ can be clearly depicted by choosing a suitable value for $s$. Since $s$ is positive, no new fixed points are introduced, and neither will it change the properties of the fixed points from the ones known previously in the hybrid variables. Its only effect is to ``stretch out'' or ``squash'' the vertical axis near the point $A$, and enable the optimal visual presentation of the dynamics. 
While the dimensionful parameter $M$ still completely cancels out of the dynamical system, the dimensionless parameter $s$ enters the equations. It does not introduce a new overall energy scale into the original system, though, but scales the dimensionless speed variable $\hat{z}$ in comparison to the dimensionless field variable $\phi$ by increasing/decreasing their ratio as an artificial ``slow/fast motion'' device in the depiction of the late stages of the evolution.

Taking the modified hybrid variables \eqref{eq: modified hybrid variables} and the hybrid time with an extra scaling factor
\eqref{eq:dttilde and dt} the dynamical system of the Higgs model \eqref{eq: H definition Higgs direct cosmic} is given by
\begin{subequations}
\label{eq: metric Higgs dyn sys modified hybrid cosmic finite}
\begin{align}
    \frac{\DD \phi}{\DD \tilde{t}} =& \frac{\hat{z}\left(H+Ms\right)}{H+M} \,, \\
    \frac{\DD \hat{z}}{\DD \tilde{t}} =& - \frac{H^{2} \hat{z} \left(36 \phi^{4} \xi^{3} + 6 \phi^{4} \xi^{2} + 48 \phi^{3} \hat{z} \xi^{3} + 8 \phi^{3} \hat{z} \xi^{2} - 6 \phi^{2} \hat{z}^{2} \xi^{2} - \phi^{2} \hat{z}^{2} \xi + 36 \phi^{2} \xi^{2} + 12 \phi^{2} \xi + 24 \phi \hat{z} \xi^{2} + 8 \phi \hat{z} \xi - 4 \hat{z}^{2} \xi - \hat{z}^{2} + 6\right)}{4 \left(H + M\right) \left(H + M s\right) \left(\phi^{2} \xi + 1\right) \left(6 \phi^{2} \xi^{2} + \phi^{2} \xi + 1\right)} \nonumber \\ 
    & - \frac{H M \hat{z} s \left(36 \phi^{4} \xi^{3} + 6 \phi^{4} \xi^{2} + 36 \phi^{3} \hat{z} \xi^{3} + 6 \phi^{3} \hat{z} \xi^{2} - 6 \phi^{2} \hat{z}^{2} \xi^{2} - \phi^{2} \hat{z}^{2} \xi + 36 \phi^{2} \xi^{2} + 12 \phi^{2} \xi + 24 \phi \hat{z} \xi^{2} + 6 \phi \hat{z} \xi - 4 \hat{z}^{2} \xi - \hat{z}^{2} + 6\right)}{2 \left(H + M\right) \left(H + M s\right) \left(\phi^{2} \xi + 1\right) \left(6 \phi^{2} \xi^{2} + \phi^{2} \xi + 1\right)} \nonumber \\
    & - \frac{M^{2} \left(6 \lambda \phi^{6} \hat{z} \xi^{2} + \lambda \phi^{6} \hat{z} \xi + 8 \lambda \phi^{5} \xi + 8 \lambda \phi^{4} \hat{z} \xi + \lambda \phi^{4} \hat{z} + 8 \lambda \phi^{3}\right)}{8 \left(H + M\right) \left(H + M s\right) \left(\phi^{2} \xi + 1\right) \left(6 \phi^{2} \xi^{2} + \phi^{2} \xi + 1\right)} \nonumber \\
    & - \frac{M^2 s^{2} \left(48 \phi^{3} \hat{z}^{2} \xi^{3} + 8 \phi^{3} \hat{z}^{2} \xi^{2} - 12 \phi^{2} \hat{z}^{3} \xi^{2} - 2 \phi^{2} \hat{z}^{3} \xi + 48 \phi \hat{z}^{2} \xi^{2} + 8 \phi \hat{z}^{2} \xi - 8 \hat{z}^{3} \xi - 2 \hat{z}^{3}\right)}{8 \left(H + M\right) \left(H + M s\right) \left(\phi^{2} \xi + 1\right) \left(6 \phi^{2} \xi^{2} + \phi^{2} \xi + 1\right)}\,,
\end{align}
where
\begin{align}
    H =& \frac{M \left(-12 \phi \hat{z} \xi s + 2 \hat{z}^{2} s + \sqrt{12 \lambda \phi^{6} \xi + 24 \lambda \phi^{5} \hat{z} \xi - 2 \lambda \phi^{4} \hat{z}^{2} + 12 \lambda \phi^{4} + 144 \phi^{2} \hat{z}^{2} \xi^{2} s^2 + 24 \phi^{2} \hat{z}^{2} \xi s^2 + 24 \hat{z}^{2} s^2}\right)}{2 \left(6 \phi^{2} \xi + 12 \phi \tilde{z} \xi - \tilde{z}^{2} + 6\right)} \,.
\end{align}
\end{subequations}
The addition of s into the definition of the hybrid variables \eqref{eq: modified hybrid variables} but keeping the time \eqref{eq:dttilde and dt} as before does not alter the qualitative properties of the system, compared to \eqref{eq: metric Higgs dyn sys hybrid cosmic finite}. When all terms are combined and factorized in \eqref{eq: metric Higgs dyn sys modified hybrid cosmic finite} the constant $M$ cancels out and the system is still explicitly dimensionless. The fixed point $A$ is present and regular at the origin, with the properties as given in Table \ref{tab: Higgs fixed points}, inherited from the fact that for negligible $H$ the hybrid variable $\hat{z}$ \eqref{eq: modified hybrid variables} is proportional to the direct scalar field variable $\bar{z}$ \eqref{eq: original variables}, with $s$ being the factor of proportionality. For large $H$ the effect of $s$ becomes insignificant, the modified hybrid variables \eqref{eq: modified hybrid variables} reduce to the Hubble rescaled variables \eqref{eq: rescaled variables}, and the equations are finite in the scalar field asymptotics as $\frac{\DD \hat{z}}{\DD \tilde{t}} \sim \phi^0$. Thus, we recover the asymptotic fixed points with the properties given in  Table \ref{tab: Higgs fixed points} without problems. In summary, the dynamical system in the modified hybrid variables \eqref{eq: modified hybrid variables} and hybrid time \eqref{eq:dttilde and dt} is globally finite and renders both the initial and final states as regular fixed points, as we have desired.

The main benefit of adding the rescaling $s$ into the definition of the variables \eqref{eq: modified hybrid variables} becomes apparent when we convert them into the  Poincar\'e form \eqref{eq: Poincare variables}, write the system \eqref{eq: metric Higgs dyn sys hybrid cosmic finite} in the global representation, and plot the respective phase portrait on Fig.\ \ref{fig: Higgs rescaled variables cosmic xi1}. The main difference compared to Fig.\ \ref{fig: Higgs hybrid variables hybrid time xi1} of the plain hybrid variables is that now the damped oscillations around the final point $A$ are obvious and easy to see. While before the ratio of $\tilde{z}/\phi$ was very small and the oscillating trajectories were ``squashed'' to the ``equatorial'' line, the rescaling parameter can adjust the ratio $\hat{z}/\phi$ to be bigger, making the oscillations more ``round'' and easier to track. In principle the parameter $s$ can take any positive value, but for the best visual effect it should be chosen such that the ratio $\hat{z}/\phi$ on the trajectories near $A$ is roughly unity. The precise value of $s$ obviously depends on the model functions and the values of the model parameters, hence $s$ can be adjusted for each particular plot one is going to make.

Besides this visual effect, the parameter $s$ does not change the other qualitative features of the system or the phase diagram. With the variables \eqref{eq: modified hybrid variables} the full global expressions get even more lengthy and uninviting, not worthy of explicit presentation here. Nevertheless, utilizing the same methods as before, one can still confirm the existence and properties of the fixed points as reported in Table \ref{tab: Higgs fixed points}. The unphysical region of the phase space is still present, as the range of $\phi, \hat{z}$ which does not correspond to real $H$. Thus overall, the hybrid variables with a rescaling factor \eqref{eq: modified hybrid variables} and the corresponding phase portrait \ref{fig: Higgs rescaled variables cosmic xi1} satisfy all the requirements we wanted to achieve. All fixed points are regular, and the dynamics is clearly discernible.

\section{Examples in nonsingular variables}
\label{sec:examples}

Having established a set of variables \eqref{eq: modified hybrid variables}, \eqref{eq:dttilde and dt} that perform very well in the case of Higgs inflation \eqref{eq: Higgs metric model}, we need to confirm that these variables will do a good job also for the other models. For this purpose let us take various example models from Ref.\ \cite{Jarv:2024krk}, and generate the global phase portraits for the same parameter values as in \cite{Jarv:2024krk}, but adopting the variables  \eqref{eq: modified hybrid variables}, \eqref{eq:dttilde and dt} instead. The parameter values in Ref.\ \cite{Jarv:2024krk} were chosen specifically to yield identical observational predictions of the scalar spectral index $n_s$, tensor-to-scalar ratio $r$, and the amplitude of the scalar power spectrum $A_s$ from the different models, close to the ones best supported by the cosmological data. We will skip the intermediate calculational expressions as these are generally quite long, while obtaining them follows exactly the same routine as in the previous section. We will just present the global phase portraits and discuss the salient features of the dynamics they depict.

\subsection{Metric Higgs inflation}

\begin{figure}[t]
	\centering
    \subfigure[]{	\includegraphics[width=0.46\textwidth]{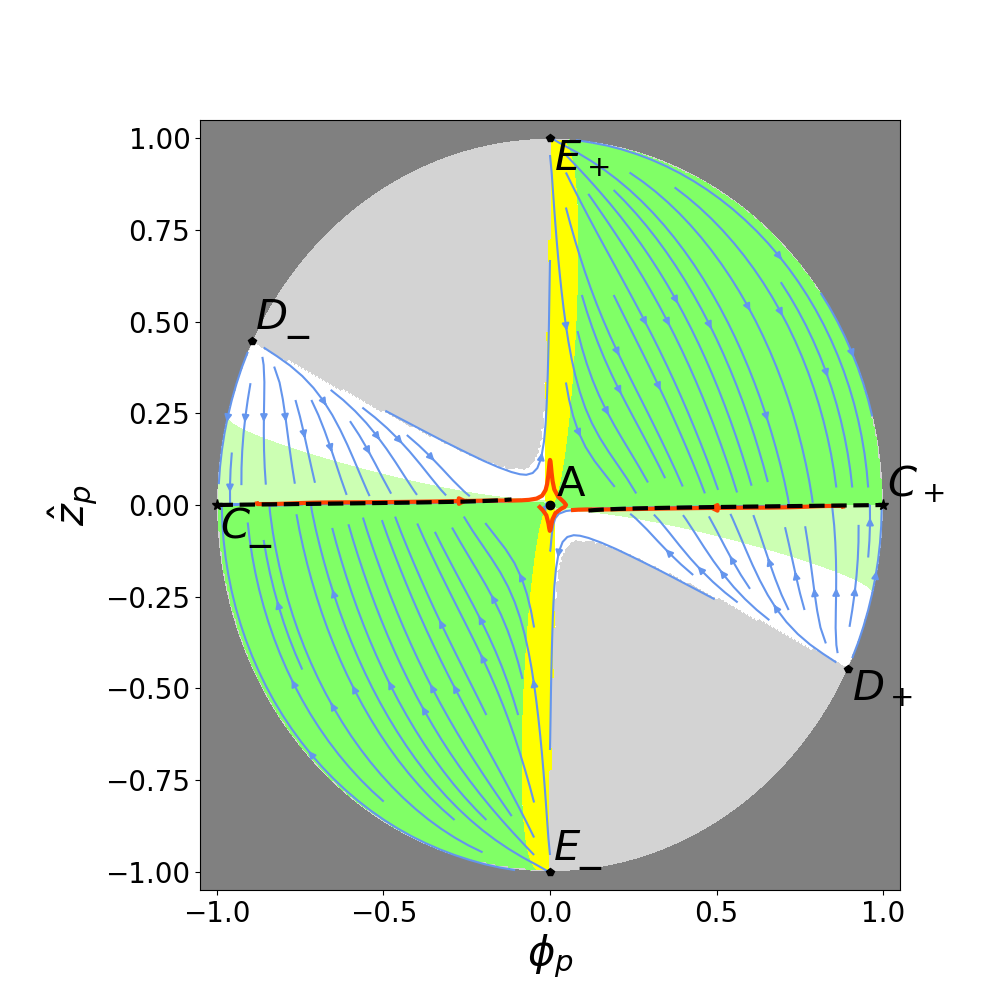} \label{fig: Higgs modified hybrid variables conformal xi100}}
	\subfigure[]{
		\includegraphics[width=0.46\textwidth]{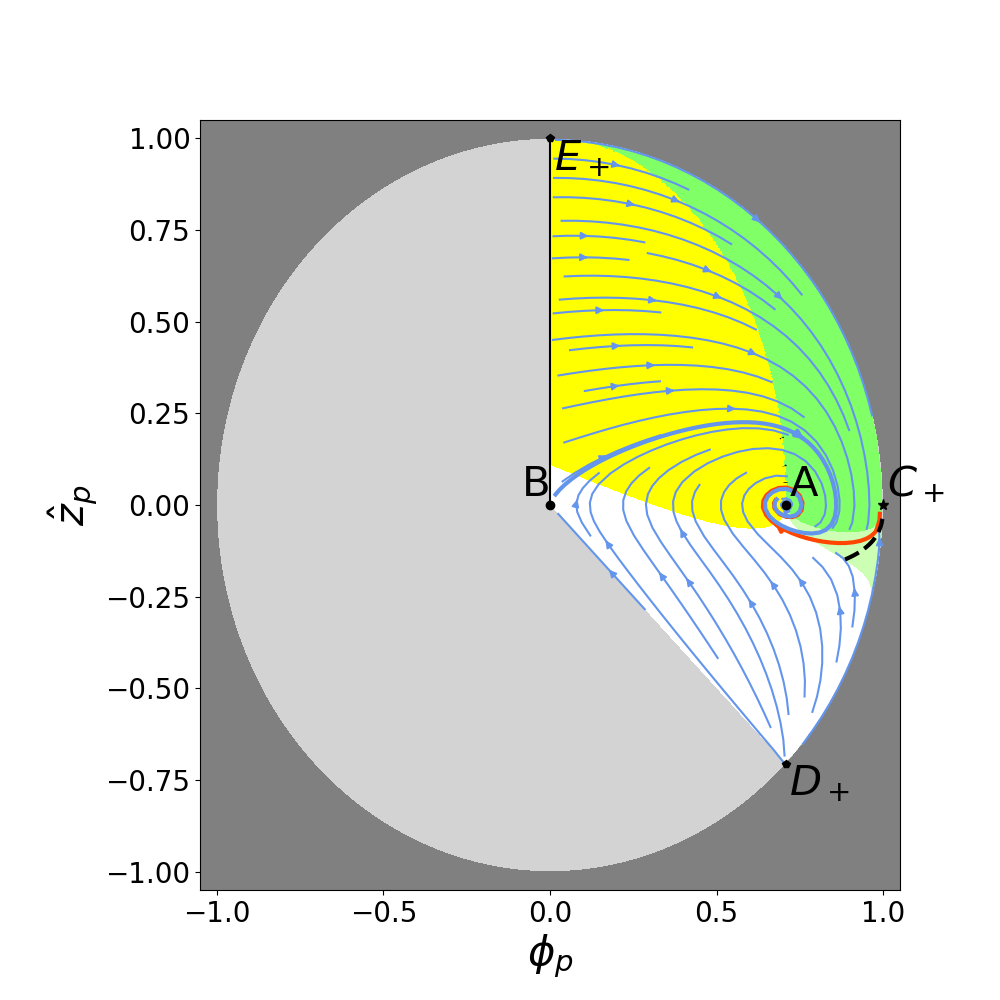} \label{fig: Starobinsky}}
   \subfigure[]{	\includegraphics[width=0.46\textwidth]{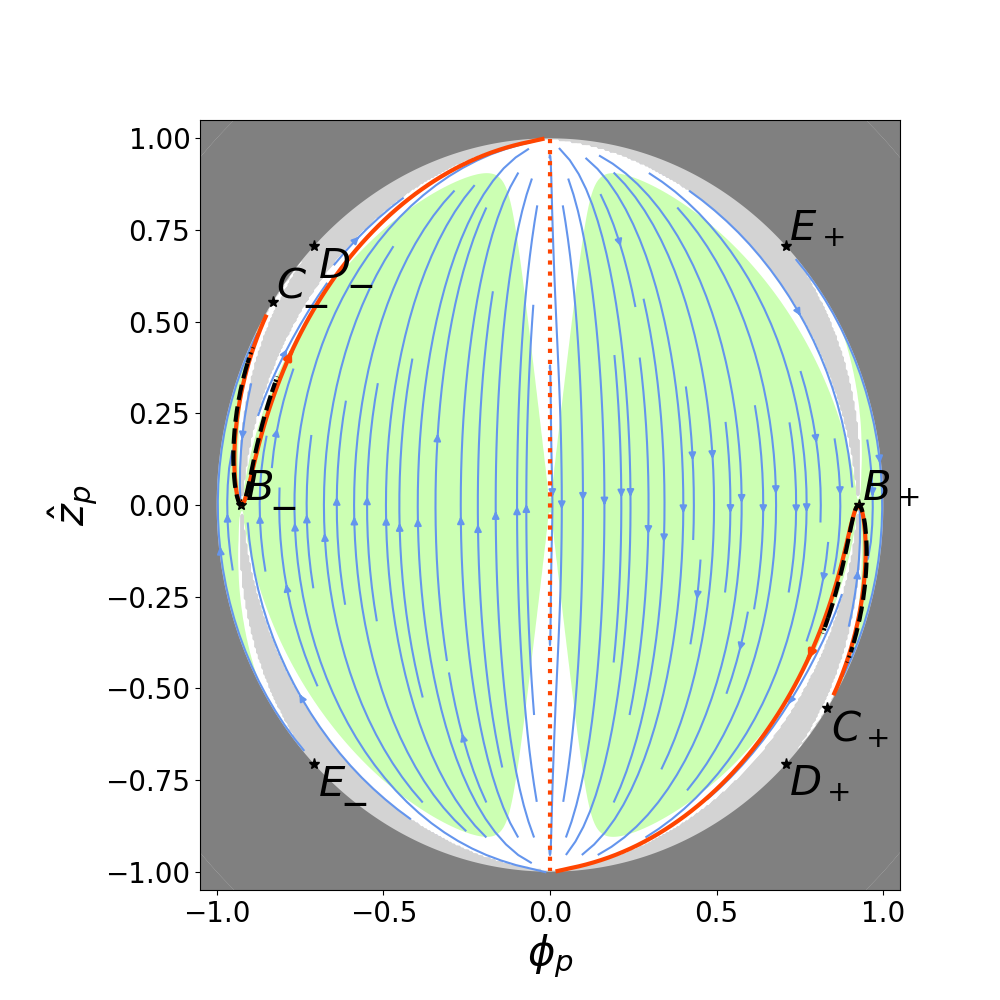} \label{fig: Pole}}
    \subfigure[]{\includegraphics[width=0.46\textwidth]{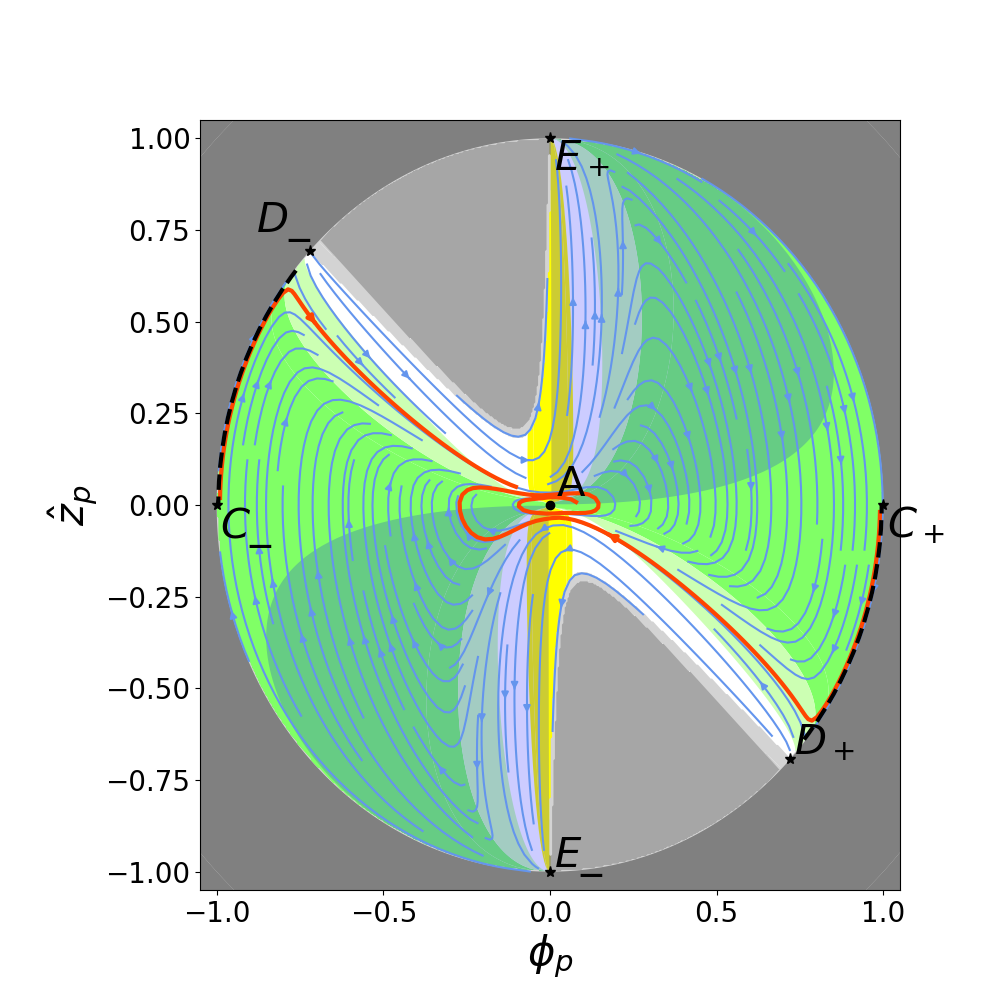} \label{fig: Nonminimal Palatini}}
  \caption{Cosmological phase portraits of different inflationary models in hybrid variables with adjustable scaling factor \eqref{eq: modified hybrid variables} in hybrid time \eqref{eq:dtt and dt}: a) the metric Higgs model \eqref{eq: Higgs metric model} with $\lambda=0.129$, $v=0$, $\xi=100$, and $s=0.005$, b) metric Starobinsky model \eqref{eq: Starobinsky metric model} with $\beta=40000$, and $s=0.005$, c) pole inflation model \eqref{eq: Pole metric model} with $\alpha=1$ and $\lambda=10^{-4}$, and $s=0.005$ d) nonminimal Palatini model \eqref{eq: nonminimal Palatini model} with $\alpha=\sqrt{2/3}$, $\lambda=10^{-5}$, $\xi=100$, and $s=0.05$. The color coding is the same as on Fig.\ \ref{fig: singular plots}, but in addition the semi-transparent blue shade designates contracting phase of the universe.  }
\label{fig: example nonsigular plots}
\end{figure}

The first example is the metric Higgs model \cite{Bezrukov:2007ep} we already encountered and analyzed before. The model functions are given by Eq.\ \eqref{eq: Higgs metric model} but the parameter values on the Fig.\ \ref{fig: Higgs modified hybrid variables conformal xi100} are different and more close to the observationally preferred ones. On the plot the rescaling parameter $s=0.005$ to make the final oscillations look as ``round'' as possible.

\subsection{Metric Starobinsky inflation}

The well known Starobinsky model \cite{Starobinsky:1980te} is  given by the Lagrangian density
\begin{align}
\label{eq: Starobinsky lagrangian}
    \mathcal{L}_{f(R)} &= f(R) = M^2 R + \beta R^2,
\end{align}
where the parameter $\beta$ is dimensionless and $M$ is  an arbitrary mass parameter, which can be selected in a way that the theory reduces to Einstein gravity at the low curvature limit. 
The expression \eqref{eq: Starobinsky lagrangian} can be written in the scalar-tensor form by introducing a scalar field $M\Phi= M\frac{\DD  f(R)}{\DD R}=M^2+2\beta R$. Such transformation yields a dynamically equivalent Lagrangian density that is akin to a non-minimally coupled scalar-tensor theory \cite{Jarv:2024krk}
\begin{align}
    \mathcal{L}_{\Phi} &= M \Phi R - \frac{M^2}{4 \beta } \left( \Phi - M \right)^2 \,.
\end{align}
From here we can immediately read off the model functions
\begin{align}
\label{eq: Starobinsky metric model}
    \mathcal{A} &= M \Phi \,, \qquad
    \mathcal{B} = 0 \,, \qquad
    \mathcal{V} = \frac{M^2}{8 \beta } \left( \Phi - M \right)^2 \,.
\end{align}
Here we assume that $\Phi$ is positive to maintain the correct sign of the effective gravitational constant. Similar stability considerations \eqref{eq: assumptions on the model functions} suggest keeping $\beta$ positive as well. 

Taking the variables \eqref{eq: modified hybrid variables}, computing their derivatives in the hybrid time \eqref{eq:dttilde and dt}, and then using the cosmological equations \eqref{eq:FLRW equations} with the model functions \eqref{eq: Starobinsky metric model} substituted in, yields a well behaved dynamical system. The respective  global phase portrait can be found on Fig.\ \ref{fig: Starobinsky}. As could be expected, the system has one regular attractive fixed point $A$ at $\phi=1$, corresponding to the minimum of the (effective) potential at $\Phi=M$. There is another fixed point $B$ of saddle type at $\phi=0$. The latter resides at the boundary of the phase space, but the dynamical system is still finite there. On the other side, it is also good that in the $\phi$ asymptotics the system remains finite and we can recover the relevant fixed points $C_+$, $D_+$, and $E_+$ in the global Poincar\'e rescaled variables. The asymptotic de Sitter point $C_+$ has saddle features and supplies a heteroclinic orbit into the point $A$, which functions as the inflationary master trajectory approximated by slow roll. The other heteroclinic orbit from $B$ to $A$ starts in a deceleration regime, and although it later experiences periods of (super)accelerated expansion, it can not account for inflation, since the slow roll regime is missing. Curiously, the line $\phi=0$ between the points $B$ and $E_+$ is not a trajectory itself, rather all points on this line correspond to specific initial conditions for different solutions. Some of these solutions eventually reach the vicinity of the inflationary master trajectory and yield sufficient inflation, as confirmed by dedicated numerical studies \cite{Mishra:2018dtg,Mishra:2019ymr}. This brings out the clear advantage of the present hybrid variables \eqref{eq: modified hybrid variables} as they make it immediately obvious which trajectories can truly approach the slow roll zone in the phase space. The Hubble rescaled variables \eqref{eq: rescaled variables} which for the same model gave Fig.\ 2a in Ref.\ \cite{Jarv:2024krk} are singular at the point $A$ and leave this aspect rather obscure.

\subsection{Pole inflation}

Our next example is pole inflation which involves the introduction of a pole in the kinetic term of the scalar field Lagrangian \cite{Terada:2016nqg}. The popular $\alpha$-attractors feature a pole of order 2 \cite{Kallosh:2013yoa, Kallosh:2015lwa, Carrasco:2015pla}, but here we focus on the model
\begin{align}
\label{eq: Pole metric model}
    \mathcal{A} &= M^2 \,, \qquad
    \mathcal{B} = \frac{6 \alpha\, M^2 \, \Phi^2}{(6\alpha M^2- \Phi^2)^2} \,, \qquad
    \mathcal{V} = \frac{\lambda}{4} \Phi^4 \,,
\end{align}
which predicts the same spectrum of of perturbations as the metric Higgs and Starobinsky models \cite{Jarv:2024krk}.
Usually in the pole inflation literature the parameter $\alpha$ has mass dimension two,
but here we have separated out the dimensionful constant $M$ to make $\alpha$ dimensionless. It is natural to set $M=\MP$. In this parametrization the poles are located at $\Phi = \pm \sqrt{6\alpha} M$. They correspond to the fixed points where the effective mass \eqref{eq: m_eff} diverges. 

The global phase portrait of the model is depicted on Fig.\ \ref{fig: Pole}. The slow roll of inflationary trajectory starts at the finite de Sitter fixed point $B$ which resides at the pole of the model. It flows towards the smaller absolute values of $\phi$ and correctly exits the accelerated expansion regime to deceleration. However, in this model the usual final fixed point $A$ is missing and the dynamical system is singular at $\phi=0$. The reason is that in the limit $\Phi \to 0$ the coupling function $\mathcal{B}$ \eqref{eq: Pole metric model} vanishes, the effective mass \eqref{eq: m_eff} turns to zero, and the scalar field equation \eqref{mpf:scalar:field:equation:friedmann:1} becomes singular. On the plot \ref{fig: Pole} this property is represented by a vertical dotted line. Contrary to a similar feature of Fig.\ \ref{fig: Higgs rescaled variables conformal xi1}, it is not the shortcoming of the dynamical variables, but stems from the model itself. The model itself is problematic, and the variables can not be blamed for that. Rather the opposite, here the hybrid variables \eqref{eq: modified hybrid variables} reveal the physical singularity, while from the simple rescaled variables \eqref{eq: rescaled variables} we would not have immediately known, whether the singularity has its origin in the choice of the variables or in the model. In addition, there are also secondary slow roll trajectories flowing between the points $C$ to $B$, but in this case there is no exit from inflation and the accelerated expansion lasts all the way to the point $B$. There is no crossing from the outer to the inner phase space region over the point $B$.

\subsection{Nonminimal Palatini model}
\label{nonminPalatini}
Finally, let us examine an example designed in the Palatini formalism and given by the model functions \cite{Jarv:2024krk} 
\begin{align}
\label{eq: nonminimal Palatini model}
\mathcal{A}&=M^2+ \xi \Phi^2 \,, \qquad 
\mathcal{B}=1 \,, \qquad 
\mathcal{V}=\lambda [M^2+\xi\Phi^2]^2 \left[ 1-
\left(\frac{M}{\sqrt{M^2+ \xi  \Phi ^2  } +  \sqrt{\xi}  |\Phi | } \right)^{\frac{\alpha}{\sqrt{\xi}}}
\, \right],
\end{align}
where $M$ again carries the dimension of mass, while the parameters $\xi$, $\lambda$, and $\alpha$ are positive and dimensionless. The potential is cumbersome but was constructed to match exactly the observational predictions of the nonminimal Higgs and Starobinsky models. The effective potential \eqref{eq: Veff} has a minimum at $\Phi=0$ which is an attractor for the late universe, and we can set $M=\MP$.

The corresponding global phase portrait is given on Fig.\ \ref{fig: Nonminimal Palatini}. In terms of the fixed point structure, unphysical regions, and generic phase flow the diagram is qualitatively similar to the metric Higgs model on Fig.\ \ref{fig: Higgs hybrid variables hybrid time xi1} and \ref{fig: Higgs modified hybrid variables conformal xi100}, while the slow roll dynamics is pushed to take place at very high values of the scalar field. The main difference is that in the Palatini case the ``$+$'' sheet of $H(\phi,\hat{z})$ that we plot here, can also feature contracting behavior of the universe, as explained after Eq.\ \eqref{eq: H in terms of Phi} and again before Eq.\ \eqref{eq: Poincare variables}. On the diagram \ref{fig: Nonminimal Palatini} the contracting phase is depiced by the semi-transparent blue shade, and we can notice how trajectories enter the contracting phase in the superstiff ($\weff>1$, yellow) region and exit back into expansion in the superaccelerated ($\weff<-1$, dark green) region. In fact, the phase portrait tells that in approaching the late universe all trajectories will periodically experience alternating contracting and expanding phases. Thus, unless the effects of reheating drastically change this feature, the current model does probably not offer a realistic scenario to describe our Universe.

Our plot \ref{fig: Nonminimal Palatini} of the nonminimal Palatini model \eqref{eq: nonminimal Palatini model} in the hybrid variables \eqref{eq: modified hybrid variables} and hybrid time \eqref{eq:dttilde and dt} can be compared to Fig.\ 2d of Ref.\ \cite{Jarv:2024krk} which depicts the same model with the same parameter values, but just using the Hubble rescaled variables \eqref{eq: rescaled variables} and e-folds time \eqref{eq:dtt and dt}. Besides the obvious aspect that the latter is singular at the origin, an attentive reader may also notice how the plot in \cite{Jarv:2024krk} features two extra ``pockets'' or regions full of trajectories in the phase space. Although these areas still belong to the real part of the ``$+$'' branch of the Friedmann constraint $H(\phi,\hat{z})$, in the present work we have excluded those regions from the physically interesting phase space, since they feature contracting behaviour and are discontinuous from the region which inhabits the slow roll trajectory. The solutions in those ``pockets'' do not enjoy the privilege of being attracted to the slow roll inflationary regime, and thus are not really relevant from the physical point of view. 

In summary, it is clear that the variables \eqref{eq: modified hybrid variables} and the measure of evolution \eqref{eq:dttilde and dt} do a good job in bringing out the relevant features of the selection of models studied in this section.
 
\section{Conclusion}
\label{sec:conclusion}

In this paper we considered flat FLRW cosmologies in metric and Palatini scalar-tensor gravity, which provide an observationally favored framework of single scalar field models for inflation. Our goal was to find a set of dynamical variables and a measure of evolution that would represent the global the phase space of the system in an optimal way. Indeed, the constructed hybrid variables with an extra adjustable scaling parameter \eqref{eq: modified hybrid variables} together with a hybrid time measure \eqref{eq:dttilde and dt} are able to describe the system in a regular manner over the entire span of the scalar field values. They allow the analysis of both the finite and infinite fixed points, and manifestly distinguish between different asymptotic states. The resulting diagrams are intuitively easy to read as the scalar field zero is located in the center, while the infinite scalar field values are mapped to the outer boundary of the compact plot. We checked that these variables work for conceptually different models and thus allow an easy comparison of the respective phase portraits next to each other. This is explicitly illustrated on Fig.\ \ref{fig: example nonsigular plots} which shows the phase portraits of the metric Higgs model, metric Starobinsky model, a pole inflation model, and a nonminimal Palatini model, for parameter choices that yield the same predictions for the perturbation observables.

Although the phase portrait of an inflationary model encapsulates the main qualitative characteristics of the background evolution only,  there is an important indirect relation to the observationally more interesting subject of perturbations as well. Namely, the observable perturbations presume the existence of a slow roll regime for the scalar field, but the latter depends on the suitable fixed point structure of the phase space. If the phase space is endowed with a saddle type and attractor type fixed points, there will be a trajectory running from the former to the latter, which draws all nearby trajectories close to itself and sustains a long period of nearly exponential spatial expansion, making inflation a generic feature of the solutions in the model. In many cases the saddle type point resides in the scalar field asymptotics, and can be clearly spotted only on the global diagram. Therefore working out the global phase portrait can give first information whether the slow roll regime is available in a model, and hence whether the calculation of perturbations is worth the effort at all. The presence of the slow roll trajectory is not automatically guaranteed, as for instance the saddle type fixed point might situate itself in the unphysical part of the phase space, ruining the prospects for inflation \cite{Jarv:2021ehj}.

Until now, we have tested the nonsingular variables in several slow roll models in scalar-tensor family of theories. Certainly, this does not exhaust the repertoire of various inflationary models, and it would be interesting to get a better phase space understanding of the fast roll, rapid roll, ultra-slow roll, constant roll, or otherwise non-slow-roll scenarios as well. In addition, it remains to be seen whether these variables will also work or can be adjusted to the cases when the scalar inflaton field is nonminimally coupled to other gravitational terms like Gauss-Bonnet or other higher curvature corrections, or as in the Horndeski theory and beyond. Finally, we may also investigate how the inflation works and what new features of the phase space occur, if the underlying geometry of the theory is extended into the teleparallel or metric-affine framework.

\section*{Acknowledgments}

This work was supported by the Estonian Research Council via the Center of Excellence ``Foundations of the Universe'' TK202U4 and the grant PRG2608.

\bibliographystyle{utphys}
\bibliography{scalartensor}

\end{document}